\documentclass[epsfig]{mn2e}
\usepackage{epsfig}

\newcommand{\ratpol}{($\frac{\Delta \lambda(\rm Pol)}{\Delta \lambda(I)}$)}
\newcommand{\av}{$A_{V}$} 

\newcommand{\ang}{$\rm \AA$}

\newcommand{\lsun}{L$_{\odot}$}
\newcommand{\msun}{M$_{\odot}$}

\newcommand{\degree}{$^{\rm o}$}

\newcommand{\ea}{{et al.}}

\newcommand{\ha}{H$\alpha$}

\newcommand{\rin}{R_{\rm in}}
\newcommand{\Rin}{R_{\rm in}}
\newcommand{\Rstar}{R_{*}}
\newcommand{\rstar}{R_{*}}

\def\kms{\,km~s$^{-1}$}      % note leading thinspace

\def\lesssim{\mathrel{\hbox{\rlap{\hbox{\lower4pt\hbox{$\sim$}}}\hbox{$<$}}}}
\def\gtrsim{\mathrel{\hbox{\rlap{\hbox{\lower4pt\hbox{$\sim$}}}\hbox{$>$}}}}

\def\arcsec{\hbox{$^{\prime\prime}$}}

\def\micron{\hbox{$\mu$m}}

\def\ion#1#2{#1$\;${\small\rm\@Roman{#2}}\relax}
 
\newbox\grsign \setbox\grsign=\hbox{$>$} \newdimen\grdimen 
\grdimen=\ht\grsign
\newbox\simlessbox \newbox\simgreatbox
\setbox\simgreatbox=\hbox{\raise.5ex\hbox{$>$}\llap
     {\lower.5ex\hbox{$\sim$}}}\ht1=\grdimen\dp1=0pt
\setbox\simlessbox=\hbox{\raise.5ex\hbox{$<$}\llap
     {\lower.5ex\hbox{$\sim$}}}\ht2=\grdimen\dp2=0pt

\makeatletter 
\renewcommand\@biblabel[1]{}     % Arabic numbers, 

\makeatother

\begin{document}

\title[Probing the circumstellar structures of T Tauri stars]
{Probing the circumstellar structures of T Tauri stars and their relationship to those of Herbig stars}

\author[Jorick S. Vink \ea ]
%{}
{Jorick S. Vink$^1$, Janet E. Drew$^1$, Tim J. Harries$^2$, Ren\'e D. Oudmaijer$^3$, Yvonne Unruh$^1$\\
$^1$Imperial College of Science, Technology and Medicine,
Blackett Laboratory, Prince Consort Road, London, SW7 2BZ, UK\\   
$^2$ School of Physics, University of Exeter, Stocker Road, Exeter EX4 4QL, UK\\
$^3$ The School of Physics and Astronomy, E C Stoner Building, Leeds, LS2 9JT, UK\\ 
}

\date{received,  accepted}

\maketitle
\begin{abstract}

We present \ha\ spectropolarimetry observations of a sample of 10 bright 
T Tauri stars, supplemented with new Herbig Ae/Be star data.
A change in the linear polarisation across \ha\ is detected in most 
of the T~Tauri (9/10) and Herbig Ae (9/11) objects, which we 
interpret in terms of a compact source of line photons 
that is scattered off a rotating accretion disk.
%For some objects, we find an increase in the degree
%of polarisation across P Cygni absorption, which yields an 
%accurate direction of the intrinsic position angle (PA) of the polarisation. This 
%``McLean'' effect, detected in FU~Ori, amongst other extreme objects, 
%may become a powerful tool for understanding disk winds in young 
%stellar objects. 
We find consistency between the position angle (PA) of the 
polarisation and those of imaged disk PAs from infrared and millimetre imaging 
and interferometry studies, probing much larger scales. 
For the Herbig Ae stars AB~Aur, MWC~480 and CQ~Tau, 
%as well as the borderline object RY~Tau, 
we find the polarisation PA to be perpendicular to the imaged 
disk, which is expected for single scattering. On the other hand, the 
polarisation PA aligns with the outer disk PA for the T Tauri stars DR~Tau and SU~Aur and  
FU~Ori, conforming to the case of multiple scattering. This difference can be explained 
if the inner disks of Herbig Ae stars are optically thin, whilst those around our T Tauri 
stars and FU~Ori are optically thick. 
Furthermore, we develop a novel technique that combines known inclination angles and our 
recent Monte Carlo models to constrain the inner rim sizes of SU~Aur, GW~Ori, AB~Aur, and 
CQ~Tau. Finally, we consider the connection of the inner disk structure with the orientation 
of the magnetic field in the foreground interstellar medium: for FU~Ori and DR~Tau, 
we infer an alignment of the stellar axis and the larger magnetic field direction. 
%the PAs of the intrinsic and foreground polarisation to be perpendicular. 
%This suggests an undisturbed star-forming collapse in which 
%the orientation of the large scale magnetic field is preserved.

\end{abstract}


\section{Introduction}

We present \ha\ spectropolarimetry of T Tauri stars as a follow-up study 
of  our earlier paper on the circumstellar structure of Herbig Ae/Be stars
(Vink et al. 2002). Spectropolarimetry is a powerful tool to probe the inner regions 
around pre-main sequence (PMS) stars. Quantifying the geometry and optical depth of the 
gaseous inner regions of such objects is crucial for studies of planet formation (Monnier et al. 2005).

It is generally believed that low-mass stars form through the collapse of 
an interstellar cloud, thereby creating a circumstellar disk. 
During the subsequent T Tauri phase, material is 
accreted from the disk onto the star, most likely 
through magnetospheric funnels 
%, given the recent evidence of magnetic fields 
(see Johns-Krull, Valenti \& Koresko 1999 and references 
therein). Whilst this basic picture of star formation is relatively well
understood, problems relating to a star's angular momentum remain, 
as we have little information on the size of the disk inner hole, the shape 
of the inflow and outflow or, in short, the geometry and kinematics 
of the circumstellar material. 
For intermediate mass (2 -- 10 $M_\odot$) Herbig Ae/Be stars, our knowledge 
becomes even more patchy (see Hartmann 1999), and for stars above 10 $M_\odot$ 
there is not even any consensus on the mode of star formation itself (see McKee \& Tan 2003).

Traditionally, the switch between low-mass and high-mass star formation
has been thought to occur at the T~Tauri/Herbig boundary 
(at $\simeq$~2 \msun), since this is where low-mass T~Tauri stars possess convective envelopes, 
whilst intermediate mass Herbig Ae star envelopes are radiative. However, recent 
data indicate that such a sharp division is no longer tenable. Herbig Ae stars
have a range of characteristics in common with T Tauri stars, varying from the
presence of inverse P~Cygni profiles, indicative of ballistic infall (Catala, Donati, \& B\"ohm 1999), 
to the detections of linear and circular line polarisations (Vink et al. 2002, Hubrig et al. 2004). These 
two types of line polarisations signal respectively the presence of compact line emission 
scattered off rotating disks, and the existence of magnetic fields for Herbig Ae stars. 
It is clear that what is needed to understand star formation as a function of mass 
(and ultimately to understand the IMF) are observations 
of the near-star environment over a wide range of PMS and other young stellar objects.
Polarimetry across emission lines is such a tool.

The application of spectropolarimetry was first established in studies of 
classical Be stars (Clarke \& McLean 1974; Poeckert 1975).  In its 
simplest form, it is based on the expectation that H$\alpha$ photons arise 
over a larger volume than the stellar continuum photons.
For this reason, the line photons scatter less frequently, and over a wider range of 
incoming angles, off the electrons in the disk. 
Hence the emission line flux will be less polarised than the continuum.
In this situation, a {\it smooth} change 
in polarisation across the line profile occurs: depolarisation.
The high incidence of such depolarisations among classical Be stars 
(26 out of 44 in Poeckert \& Marlborough 1976) indicated 
that the envelopes of classical Be stars are not spherically symmetric, and   
these findings are now taken as proof that classical 
Be stars are embedded in circumstellar disks (e.g. Dougherty \& Taylor 1992; 
Wood, Bjorkman \& Bjorkman 1997; Quirrenbach et al. 1997).  

In recent years, the technique has been applied to PMS Herbig 
Ae/Be stars (Oudmaijer \& Drew 1999; Vink et al. 2002). We have found that 
for the Herbig Be stars the frequency\footnote{Note that a disk that is observed exactly face-on 
is circular on the sky and will not yield polarisation} of line effects (7/12) 
as well as its character (the above described ``depolarisation'' effect) hint at the 
presence of undisrupted disks around these objects.
For the Herbig Ae stars, we found rather different \ha\ polarisation results. 
Here, the frequency was somewhat higher (9/11), and most intriguing was that the polarisation 
changes across \ha\ were narrower than the width of the intensity profile itself -- which is inconsistent 
with depolarisation. Instead, the data show intrinsic {\it line} polarisations that are caused by \ha\ photons from 
a compact source that scatter off an exterior medium. In contrast to the depolarisations in the Herbig 
Be stars, these line polarisations seen in Herbig Ae stars yield kinematic information.
The rotations in the position angle (PA) of the polarisation across the line profile 
translate into ``loops'' when the data are plotted in the Stokes $QU$ plane. These $QU$ loops directly 
imply the presence of envelope rotation (Wood, Brown \& Fox 1993; Vink, Harries \& Drew 2005) and suggest 
the existence of rotating accretion disks, similar to those accepted to be present in the 
lower mass T~Tauri counterparts.

Given the presumed similarities between Herbig Ae stars and T Tauri stars, 
we presented the first medium resolution 
($R$ $\simeq$ 9~000) \ha\ spectropolarimetry of a T~Tauri star: RY~Tau (Vink et al. 2003). 
The observed $QU$ loop in RY~Tau showed the presence of an inner rotating disk, which is 
consistent with the results of longer wavelength imaging studies (Koerner \& Sargent 1995) that probe the disk on larger spatial scales. In addition, 
our data provided a strong link between this classical T~Tauri star of spectral type F8, 
and the earlier-type Herbig Ae stars, where these $QU$ loops are found to be common (Vink et al. 2002). 

For the even later type (G-K) T Tauri stars narrow-band linear polarimetry 
has been performed (e.g. Bastien 1985), but no polarisation changes have been 
found to date. This could be due to a lack of sufficient 
spectral resolution in the older data. Our example of medium-resolution ($R$ $\simeq$ 9~000) line 
polarimetry on RY~Tau has shown that degrading the resolution can ``wipe out'' line 
polarisation changes (Vink et al. 2003). 
We note that the presumed absence of \ha\ polarisation features has often
been used to argue that polarisation in T~Tauri stars is simply due to scattering off dust 
-- rather than off gas. Our above-mentioned data on RY~Tau suggest that such a conclusion is 
premature, and that with the higher spectral resolution currently available, we may have to 
readdress these original deductions.

Furthermore, to be able to study the general relationship between Herbig Ae and T Tauri stars, 
it is crucial to observe more than just one T~Tau star at the relatively early spectral type F8. 
We will see that although the line polarimetry 
of T~Tauri stars harbours a variety of characteristics, the general behaviour is a predominance 
of $QU$ loops, similar to those present in Herbig Ae stars. We supplement the T~Tauri data with 
new (higher quality) Herbig Ae/Be star data, but the comparison between the T~Tauri stars presented 
here and the Herbig Ae/Be stars is mostly based on the Herbig data presented in Vink et al. (2002).

An important quantity we may be able to obtain from medium-resolution ($R$ $\simeq$ 9~000) 
spectropolarimetry across 
emission lines of PMS stars is that of the intrinsic PA of the polarisation. 
For an optically thin, single-scattering disk, 
the expected intrinsic PA of the polarisation is expected to be perpendicular to the disk PA
(e.g. Brown \& McLean 1977). Instead, for an optically thick, multiple scattering disk, 
the PA of the polarisation is expected to be parallel to the PA of the imaged 
disk (e.g. Angel 1969). 

Spectropolarimetry can also be used to set constraints on the radii of disk inner rims.
A novel technique was recently introduced by Vink et al. (2005). Their 
Monte Carlo scattering predictions show a difference in the wavelength dependence of the PA 
across emission line profiles for the case of an undisrupted disk 
versus one with an inner hole. This behaviour can be used to constrain the radii of disk 
inner holes.

Last but not least, from measured excursions in the $QU$ plane, one may 
determine the direction of polarisation in the foreground of the PMS star, and obtain information 
on the orientation of the magnetic field in the foreground medium, and thereby gain insight into 
the star formation process itself. 

The paper is organised as follows. 
In Sect.~\ref{s_obs} we discuss how the observations were obtained. 
In Sect.~\ref{s_results}, we present the general results on the continuum and line polarimetry. 
We then describe the data for the individual T~Tauri stars, in combination with the results obtained from 
other techniques (Sect.~\ref{s_ind}). 
Complementary data on Herbig Ae/Be stars are found in  
Sect.~\ref{s_herbig}. The two object classes are compared in Sect.~\ref{s_disk}, where we 
discuss the main outcomes of our study, before we conclude in Sect.~\ref{s_con}.

\begin{table*}
\begin{minipage}{\linewidth}
\renewcommand{\thefootnote}{\thempfootnote}
\caption{Targets. The {\it V}-band magnitudes (unless the {\it B}-band is specified) are taken 
from {\sc simbad}, and listed in column (2). 
The Spectral types (column 3) are taken from the Herbig \& Bell (HB) catalogue, unless they have been improved, 
when they are footnoted. 
The T Tauri type (from HB) is given in column (4).
The integration times (column 6) denote the total exposures. 
The continuum PA and its error are indicated in column (8), whilst column (9) indicates whether we have 
good \ha\ line polarisation data; the $+$ indicates there is a plot of the  epoch.
Finally, column (10) indicates a measure of the disk PA derived from line excursions. 
}
\label{t_cont}
\begin{tabular}{lclcclcrlc}
\hline
Name & {\it V} & Spec. Tp & Type of object & Date & Exposure(s) & $P_{\rm cont}^{\rm R}$ (\%) & $\Theta_{\rm cont}^{\rm R}$ (\degree) & \ha\ data? & $\Theta_{\rm intr}^{\rm R}$ \\
\smallskip\\
 (1)   &  (2)  &  (3) & (4)  &  (5)  & (6)   & (7)  & (8)   & (9)  & (10) \\
\hline
RY Tau      & 10.2 &F8$^1$& ctts&26-12-01& 4$\times$120,12$\times$180& 3.156 $\pm$ 0.007 & 14.7 $\pm$ 0.1 & yes & 163\degree\\
            &      &     &      &10-12-03& 8$\times$120,20$\times$240& 0.831 $\pm$ 0.005 & 27.9 $\pm$ 0.2 & yes & \\
            &      &     &      &12-12-03& 8$\times$90,12$\times$180 & 2.204 $\pm$ 0.004 &  9.3 $\pm$ 0.0 & yes & \\
            &      &     &      &13-12-03& 8$\times$120,12$\times$180& 2.064 $\pm$ 0.004 &  8.8 $\pm$ 0.1 & yes & \\
T Tau       &9.6$B$&K0$^2$& ctts&27-12-01& 8$\times$45,16$\times$120 & 0.663 $\pm$ 0.009 & 94.5 $\pm$ 0.4 & yes $+$ & \\
            &      &     &      &12-12-03& 4$\times$60,16$\times$150 & 1.133 $\pm$ 0.004 & 88.8 $\pm$ 0.1 & yes $+$ & \\
SU Aur      & 9.2  & G2  &su aur&27-12-01& 12$\times$180      & 0.930 $\pm$ 0.008 &104.9 $\pm$ 0.2 & yes & \\
            &      &     &      &10-12-03& 8$\times$90,16$\times$150 & 0.510 $\pm$ 0.004 & 99.0 $\pm$ 0.2 & yes & \\
            &      &     &      &11-12-03& 4$\times$90, 5$\times$240 & 0.590 $\pm$ 0.002 &102.0 $\pm$ 0.1 & yes $+$& 130\degree\\
            &      &     &      &13-12-03& 12$\times$120      & 0.512 $\pm$ 0.006 & 99.0 $\pm$ 0.3 & yes & \\
FU Ori      & 8.9  & G3  &fu ori&27-12-01 & 4$\times$20        & 0.770 $\pm$ 0.038 &132.1 $\pm$ 1.4 & no  & \\
            &      &     &      &11-12-03 & 8$\times$60        & 0.659 $\pm$ 0.009 &129.6 $\pm$ 0.4 & yes $+$ & 45\degree\\
CO Ori      &10.6  &F7$^1$&su aur&27-12-01& 8$\times$120,8$\times$180 & 2.007 $\pm$ 0.014 &168.9 $\pm$ 0.2 & yes & \\
            &      &     &      &12-12-03& 4$\times$60        & 2.506 $\pm$ 0.068 &159.2 $\pm$ 0.8 & yes $+$& (60)\degree\\
DR Tau      &13.6&K5$^1$ &ctts&27-12-01& 16$\times$180      & 0.362 $\pm$ 0.016 &142.1 $\pm$ 1.3 & yes & \\
            &      &     &      &12-12-03& 4$\times$60,20$\times$180 & 0.285 $\pm$ 0.008 &138.1 $\pm$ 0.8 & yes $+$ & 120\degree\\
RW Aur A    &10.3  &K3$^3$& ctts&10-12-03& 12$\times$150      & 0.636 $\pm$ 0.009 &103.5 $\pm$ 0.4 & yes & \\
            &      &     &      &13-12-03& 4$\times$60,20$\times$150 & 0.727 $\pm$ 0.006 &119.2 $\pm$ 0.2 & yes $+$ & 115\degree\\
GW Ori      & 9.9  &G5$^4$&ctts &11-12-03& 4$\times$60,20$\times$240 & 0.346 $\pm$ 0.003 &113.8 $\pm$ 0.3 & yes $+$ & (60)\degree \\
V410 Tau    &10.6  & K3  & wtts &12-12-03& 4$\times$90        & 0.206 $\pm$ 0.016 &153.8 $\pm$ 2.2 & no  & \\
V773 Tau    &10.7  & K3  & ctts &13-12-03& 4$\times$60,16$\times$240 & 0.321 $\pm$ 0.004 & 78.8 $\pm$ 0.3 & no  & \\
UX Tau A    &10.6$B$&K2$^5$&wtts&13-12-03& 4$\times$120,24$\times$240& 0.572 $\pm$ 0.005 & 55.5 $\pm$ 0.3 & yes $+$ & 140\degree\\
CoKu HP Tau &      &     &su aur&13-12-03& 4$\times$30        & 2.284 $\pm$ 0.029 & 65.8 $\pm$ 0.4 & no  & \\
BP Tau      &10.7$B$&K7  & ctts &13-12-03& 4$\times$30,24$\times$300 & 0.102 $\pm$ 0.006 & 70.2 $\pm$ 1.8 & yes $+$ & \\ 
\hline
\end{tabular}
\\
\noindent
$^1$ Mora et al. (2001) 
$^2$ Ghez et al. (1991)
$^3$ Rucinski (1985)
$^4$ Bouvier \& Bertout (1989) 
$^5$ Bouvier et al. (1986)
\end{minipage}
\end{table*}

\section{Observations}
\label{s_obs}

Our targets were selected from the Herbig \& Bell (1988; hereafter HB) catalogue 
on the basis of their relative brightness ($V \la 11$) and their position on the sky. 
We note that they were not chosen on the basis of known circumstellar geometries, 
or T Tauri type (i.e. classical, weak-line, naked, Fu Ori, Su Aur type, etc.).
The list of objects is given in Table~\ref{t_cont}, alongside their {\it V} magnitudes and 
spectral types.

The linear spectropolarimetry data were obtained during the 
nights of 2001 December 26 and 27, and 2003 December 10 -- 13 with 
the ISIS spectrograph on the 4.2-metre William 
Herschel Telescope (WHT), La Palma. In general, the data from the 2003 run are of a 
higher quality than those from 2001, and we therefore concentrate on those from 2003.
The dates and exposure times of all epochs are given in columns (5) and (6) of Table~1. Note 
that a majority of targets was observed on more than one occasion, to search for rotational 
modulation of spectropolarimetric signatures.
The observations were obtained in a very similar manner to that of the Herbig Ae/Be 
study by Vink et al. (2002), and we refer to that paper for details.
For the set-up specifics of the 2001 December run, we refer the reader to 
the paper on RY~Tau by Vink et al. (2003), but note that the spectral resolution
was similar to that of the 2003 data.
For the observations taken in 2003 December, we employed a slit width of $1.0 \arcsec$. 
We used the MARCONI2 CCD detector with the R1200R grating, with a spectral 
range of 1000 \ang, centred at 6500 \ang. 
This set-up corresponded to a spectral resolution
of $\simeq$ 35 \kms~around \ha\ -- as measured from arc line fits.
To be able to analyse the linearly polarised component in the spectra, 
ISIS was equipped with the appropriate polarisation optics, i.e. 
a rotating half-wave plate and a calcite block.

The data reduction was carried out using {\sc iraf}, 
and included the usual bias-subtraction, 
flat-fielding, cosmic ray removal, spectrum extraction 
and wavelength calibrations of both the ordinary and the extraordinary ray.
These were subsequently imported into the starlink package {\sc ccd2pol}. 
The Stokes parameters $Q$ and $U$ were determined, leading
to the percentage linear polarisation $P$ and its PA $\theta$:

\begin{equation}
P~=~\sqrt{(Q^2 + U^2)}
\end{equation}
\begin{equation}
\theta~=~\frac{1}{2}~\arctan(\frac{U}{Q})
\end{equation}
Note that a PA of 0\degree, i.e. North, on the sky is represented by
a vector that lies parallel to the positive $Q$ axis, whereas $\theta$\,=\,90\degree\ (i.e. 
East) is positioned in the negative $Q$ direction. Positive and negative $U$ axes 
thus correspond to position angles of respectively 45\degree\ and 135\degree.

The achieved accuracy of the polarisation data is in principle determined by 
photon-statistics only, and can be rather good (typically 0.01\%). Nonetheless, 
the quality and the amount of data taken on spectropolarimetric standard stars 
is only sufficient to reach absolute accuracies of $\simeq$ 0.1\%.

In most parts of the paper the polarisation spectra are shown binned
to a constant error of 0.10\% polarisation. Therefore, the presented 
spectra exhibit a resolution that depends on the number of counts at each wavelength. 
For this reason, the highest resolution is 
achieved at the wavelength of the \ha\ emission peak. In a few cases, we choose
to modify the error per bin.  This is done to achieve 
the best compromise between minimising the error per bin and 
resolving the line profile.

We do not correct for instrumental or interstellar polarisation, as these 
only add a wavelength-independent vector to all observed polarisations when 
plotted in the $QU$ plane. Because of this wavelength independence, we 
generally concentrate on the $QU$ representation of the data 
in classifying the type of \ha\ polarisation signature.
We note that the instrumental polarisation is extremely low ($<$ 0.1\%) with ISIS/WHT.
In addition, the interstellar polarisation between us and Taurus-Auriga 
is also low, because of the proximity to this complex. 
Hence, the only significant intervening polarisation source  
to that of the intrinsic polarisation is that of {\it foreground} polarisation 
associated with that of the molecular cloud. 
The distribution of this foreground material on the sky can be patchy, and the often-used 
method of using the polarisations of neighbouring field stars to obtain intrinsic polarisations 
may lead to dubious results in the case of young stars.
Instead, with line polarimetry, one can find the intrinsic PA angle {\it independent} 
of the foreground material. 

Occasionally, the morphology of the excursion in the $QU$ plane allows 
one to derive the direction of the foreground polarisation, and obtain the orientation 
of the environment's magnetic field. In such cases, the intrinsic PAs of the PMS are 
perpendicular to those of the foreground, which would suggest an undisturbed star-forming 
collapse, where the orientation of the environment's magnetic field is 
preserved.

\begin{table*}
\begin{minipage}{\linewidth}
\renewcommand{\thefootnote}{\thempfootnote}
\caption{The \ha\ line results. 
The errors on the equivalent widths of the \ha\ lines (column 2) 
are below 5\%, the errors on $\Delta \lambda(\rm Pol)$ (column 5) 
and $\Delta \lambda(I)$ are determined at Full Width Zero Intensity (FWZI) 
and are about 10\%.
$\Delta \lambda(\rm Pol)$ has been defined as the width over 
which the polarisation changes. In the case where the widths in PA 
and \%Pol are unequal, we take the largest of the two.  
The fractional width \ratpol\ is given in Column (6).  
The recipe with regard to the depolarisation question (column 7) 
is described in the text. Column (8) represents the morphology in $QU$ 
space. 
\label{t_line} }
\begin{tabular}{lcccccclc}
\hline
Object & \ha\ EW(\ang) & Line/Cont &  Line  & $\Delta \lambda(\rm Pol)$ & ($\frac{\Delta \lambda(\rm Pol)}{\Delta \lambda(I)}$) & depolarisation? & $QU$ & Mean\\
       &  & contrast  & effect? & (\ang) & & & behaviour & character\\
\smallskip\\
  (1) & (2)  & (3)  & (4)  & (5) & (6) & (7) & (8) & (9) \\
\hline
RY Tau      & $-$10 & 2.1  & Yes & 16  & 0.50 & No & loop   & Loop   \\
            & $-$20 & 3.2  & Yes &     &      & No & loop   &    \\
            & $-$14 & 2.6  & Yes &     &      & No & loop   &    \\
            & $-$16 & 2.8  & Yes &     &      & No & loop?  &    \\
T Tau       & $-$54 & 9.5  & Yes? & --  & --  &Yes?& --     & None\\
            & $-$35 & 10   & No  & --  & --   & -- & --     &   \\
SU Aur      & $-$6.3& 1.7  & Yes &     &      & No & exc    & Loop \\
            & $-$7.7& 2.1  & Yes & 15  & 0.60 & No & loop   &    \\
            & $-$4.6& 1.9  & Yes &     &      & No & loop   &    \\
            & $-$3.9& 1.5  & Yes &     &      & No & loop   &    \\
FU Ori      & $+$2.3& 0.3  & Yes &  9  & 0.33 & No & exc    & McLean \\
CO Ori      & $-$6.3& 2.0  & Yes &     &      & No & exc/loop?   & Exc/Loop   \\
            & $-$49 & 16   & Yes?& 19  & 0.53 & No & --     &    \\
DR Tau      & $-$33 & 12   & Yes &     &      & No & exc    & McLean\\
            & $-$57 & 15   & Yes & 30  & 0.17 & No & loop?  &    \\
RW Aur A    & $-$20 & 7.3  &  ?  & 22  & 0.50 & No &  --    & Loop/McLean \\
            & $-$70 & 11   & Yes &     &      & No & loop?  &    \\
GW Ori      & $-$27 & 7.5  & Yes & 16  & 0.69 & No & loop   & Loop   \\
UX Tau      & $-$9.0& 3.4  & Yes & 13  & 0.85 & Yes& exc    & Depol   \\
BP Tau      & $-$70 & 15   & Yes & 17  & 0.41 & No & loop?  & Loop   \\ 
\hline
\end{tabular}
\\
\end{minipage}
\end{table*}

\begin{figure}
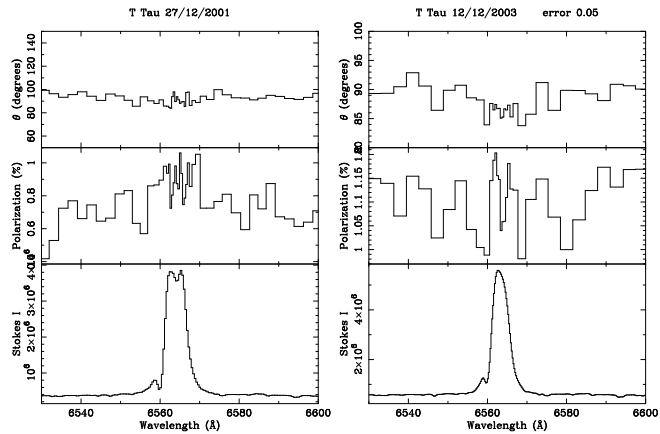

\mbox{
\epsfxsize=0.24\textwidth\epsfbox{tt.ps}
\epsfxsize=0.24\textwidth\epsfbox{tt2.ps}
}
\caption{Triplots of the prototype T Tauri star: T Tau -- 
at two different epochs.
The total light, Stokes $I$, spectrum is shown 
in the lowest panel of the triplot, 
the \%Pol is given in the middle panel, 
whilst the PA ($\theta$; see Eq.~2) is 
plotted in the upper panel. The data are rebinned such that 
the 1$\sigma$ error in the polarisation corresponds to 0.10\%, 
as calculated from photon statistics. 
In case of exceptions on this, the polarisation errors are given 
on top of the triplots. 
}
\label{f_noline}
\end{figure}

\section{Results}
\label{s_results}

We begin with a brief presentation of our continuum 
polarisation results (Sect.~\ref{s_cont}), before we focus on 
the line polarimetry (Sect.~\ref{s_spec}). 
For many objects, repeat observations are available.
Here, we concentrate on the \ha\ line polarimetry data from a single 2003
epoch.
However, we do consider historical continuum polarimetry in the discussions of individual T Tauri stars
(Section~4).

\subsection{Continuum Polarimetry}
\label{s_cont}

The linear polarisation of PMS continua can be attributed to 
the scattering of stellar photons off matter within
an asymmetric circumstellar geometry. In addition, there may be 
a foreground contribution. The polarisation of T Tauri stars 
is known to be variable in a variety of cases (e.g. Vardanyan 1964; Serkowski 1969;
Bastien \& Landstreet 1979; Hough et al. 1981; Bastien 1982;   
Schulte-Ladbeck 1983; Bastien 1985; Drissen, Bastien \& St-Louis 1989; M\'enard \& Bastien 1992; 
Grinin, Kolotolov \& Rostopchina 1995; Gullbring \& Gahm 1996; Heines, Henning \& Szeifert 1997; Yudin 2000; 
Oudmaijer et al. 2001), and it is clear that (at least part of) the polarisation is 
intrinsic to the source, as the foreground contribution is likely to be non-variable.  

In principle, the variability of the observed continuum polarisation may be used to 
identify its origin. It has been suggested that rotational modulation 
may signal cool and hot spots on the stellar surface (e.g. Stassun \& Wood 1999), but 
also variable extinction by a dusty disk can lead to quasi-periodic behaviour 
of the observed level of dust polarisation (e.g. M\'enard \& Bastien 1992).

Partly based on what Bastien (1982) noted as a general absence of polarisation changes 
across \ha\ in T~Tauri stars, he advanced what is now the commonly
accepted view on the origin of T Tauri polarisation: namely that 
it is due to scattering off extended dusty envelopes, where the geometry and kinematics 
of gas does not play a role. Whether this view represents the entire picture is a different story, as it 
has recently been challenged by the discovery of polarisation changes
across \ha\ in the classical T~Tauri star RY~Tau (Vink et al. 2003) and the results presented here. 
It is therefore not yet settled which 
polarigenic agent causes the circumstellar polarisation in T Tauri stars.
Instead, this issue should be considered open, until one can predict the polarisation of the objects, as a function of 
wavelength, through spectral lines, as a function of time.

The measured continuum polarisations are 
summarised in Table~\ref{t_cont}. The mean percentage polarisation 
and PA appear in columns (7) and (8). 
In addition, for cases where an excursion in $QU$ space is present, 
we can estimate the sky PA from these line excursions, and 
present them in column (10). We emphasise that 
the presumed disks would be expected to lie orthogonal (i.e. at 90\degree) from this position angle,  
if multiple scatterings are negligible.

\subsection{\ha\ line polarimetry}
\label{s_spec}

The observed \ha\ characteristics of each target are listed 
in Table~\ref{t_line}. The first few columns list the parameters 
as deduced from the Stokes $I$ profiles only: the equivalent 
width (EW; column 2) and the line over continuum contrast (column 3). 

\begin{figure*}
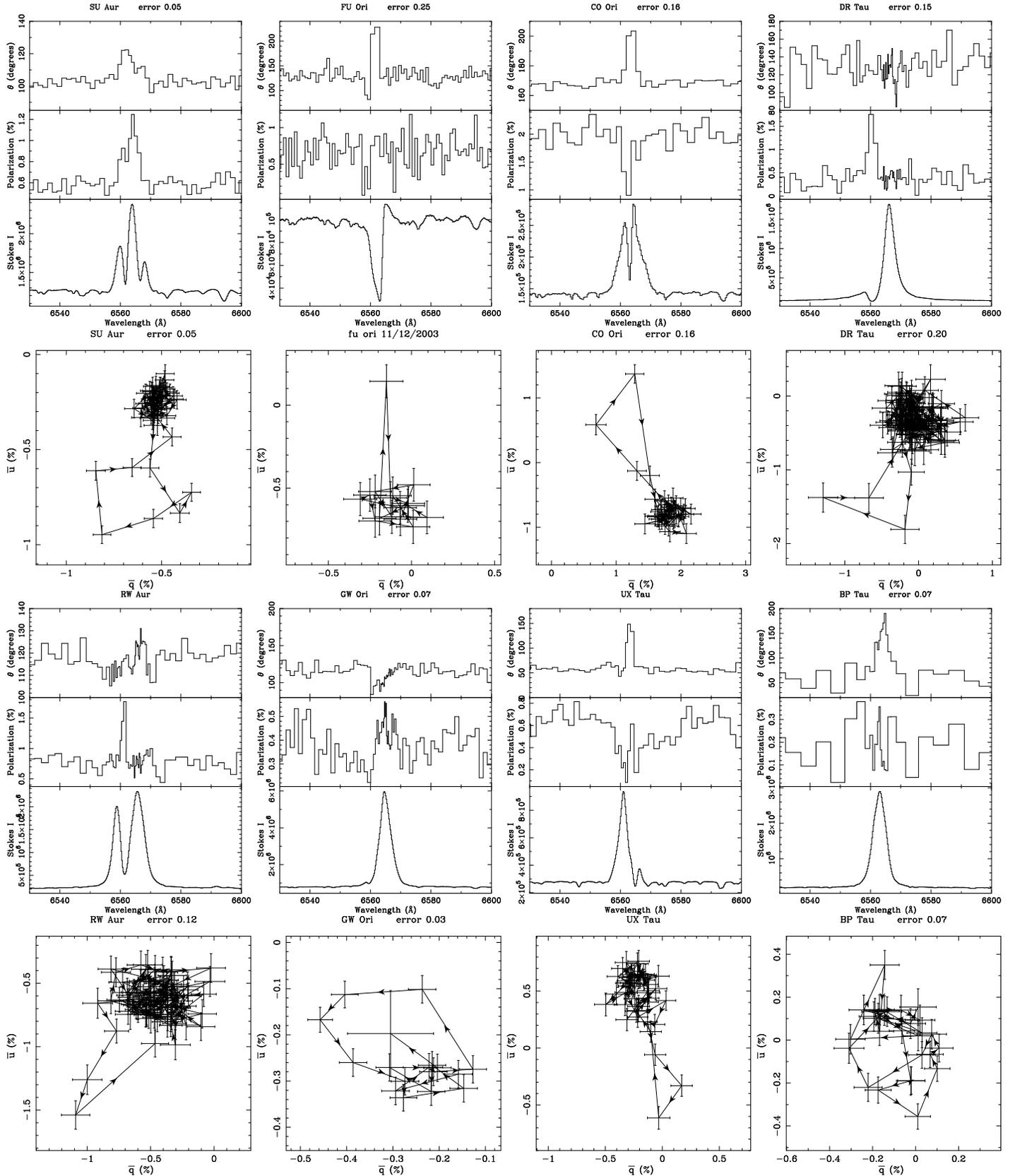

\mbox{
\epsfxsize=0.26\textwidth\epsfbox{su2.ps}
\epsfxsize=0.26\textwidth\epsfbox{fu.ps}
\epsfxsize=0.26\textwidth\epsfbox{co_o.ps}
\epsfxsize=0.26\textwidth\epsfbox{dr.ps}
}
\mbox{
\epsfxsize=0.26\textwidth\epsfbox{qu_su2.ps}
\epsfxsize=0.26\textwidth\epsfbox{qu_fu.ps}
\epsfxsize=0.26\textwidth\epsfbox{qu_co_o.ps}
\epsfxsize=0.26\textwidth\epsfbox{qu_dr.ps}
}
\mbox{
\epsfxsize=0.26\textwidth\epsfbox{rw4.ps}
\epsfxsize=0.26\textwidth\epsfbox{gw.ps}
\epsfxsize=0.26\textwidth\epsfbox{ux.ps}
\epsfxsize=0.26\textwidth\epsfbox{bp.ps}
}
\mbox{
\epsfxsize=0.26\textwidth\epsfbox{qu_rw4.ps}
\epsfxsize=0.26\textwidth\epsfbox{qu_gw.ps}
\epsfxsize=0.26\textwidth\epsfbox{qu_ux.ps}
\epsfxsize=0.26\textwidth\epsfbox{qu_bp.ps}
}
\caption{
Polarisation spectra (see Fig.~\ref{f_noline}) of the line effect T Tauri stars and their 
$QU$ diagrams (below the triplots). 
The lower plot represents the normalised Stokes parameters
$u = U/I$ against $q = Q/I$.
}
\label{f_line}
\end{figure*}

The Stokes $I$ \ha\ line shapes of the target stars vary 
from pure emission, to P Cygni and inverse P Cygni type, through 
to clear double-peaked, and even more complicated line profiles.
We have already seen in Vink et al. (2002) that Stokes $I$ line profiles 
do not necessarily correlate with the line 
polarimetry, which is here confirmed. Whilst the Stokes $I$ line profiles 
may vary dramatically from one object to the next, the different stars often still 
share similar Stokes $Q$ and $U$ profiles. This is 
illustrative of the fact that polarised light has 
a more selective origin than total light, in that intensity 
contains both direct and scattered light, whereas polarised light only contains 
the latter. 
Columns (4) to (8) in the Table concern line polarimetry properties.
We note that we have only measured the widths of the line profiles 
for the data that we actually plot. However, we do address the line 
polarimetry behaviour at all epochs (in columns 7 and 8), to make 
the best judgement for the mean polarisation character of the object under 
consideration (column 9).

Contrary to our published work on the classical T~Tauri star RY~Tau (Vink et al. 2003), 
the prototype, T~Tau itself, stands out in that it does not appear to show a significant 
line effect, cf. Fig.~\ref{f_noline}. The figure shows polarisation spectra of T~Tau from both 
2001 and 2003 presented as triplots (consisting of Stokes $I$, $P$, and $\theta$). 
The data for all other T~Tauri objects do show a significant variation 
in either the polarisation and/or the PA across the line, and these are presented in Fig.~\ref{f_line}, 
as triplots and also as loci in the $QU$ plane.
The high number of line effect detections for T Tauri stars, 9 out of 10 (including RY~Tau), is 
very similar to that found for Herbig Ae stars (9/11, Vink et al. 2002), and contradicts 
earlier narrow-band polarimetry results, which did not have the required spectral resolution.

Although the morphology in both the polarisation triplots 
and  $QU$ space shows some variety between objects, we classify them according to their common 
characteristics, and apply the same two measures as those for Herbig Ae/Be stars (Vink et al. 2002). 
One measure involves the fractional line width \ratpol\ over 
which the polarisation changes (see column 6 in Table~\ref{t_line}).
The other is the description of the shape of the 
polarisation change across the line (columns 7 and 8).  Whether or not the 
change in polarisation across \ha\ is consistent with the depolarisation 
mechanism is noted in column (7). The recipe followed to come to the 
answer is the same as that in Vink et al. (2002): if the fractional width \ratpol\
is larger than 0.75 and the behaviour in \%Pol and PA is smooth across the line, 
it is consistent with depolarisation; in case the fractional width \ratpol\ is   
less than 0.75, and there is a rotation across the line profile, causing 
a loop in the $QU$ plane (column 8), the behaviour is inconsistent with depolarisation and 
intrinsic {\it line} polarisation is inferred. 
In every case we expect to see a dark knot of 
points in the $QU$ plane, which arises from the sampling of the continuum polarisation.  
Where there is a more or less linear excursion arising from the continuum knot, we 
refer to this excursion with ``exc''. 
The mean character (column 9) combines the information from the $QU$ plane and the triplots 
of all available epochs.

\begin{table}
%\begin{minipage}{\linewidth}
\renewcommand{\thefootnote}{\thempfootnote}
\caption{Previous continuum polarimetry. Columns (2) and (3) give the number of measurements and the wavelength at which 
the polarisation was measured. Columns (4) and (4) list the ranges (not errors) of measured \%Pol and PA. The references are given in column (6) and listed below the table.} 
\label{t_earlier} 
\begin{tabular}{lccccr}
\hline
Object & Number & Lambda & $P_{\rm cont}$ & $\Theta_{\rm cont}$ & Ref\\
       & of meas      & (\AA)      &  (\%)          & (\degree) & \\
 (1)   &   (2)    & (3)   & (4)  & (5)   & (6) \\
\hline
RY Tau       &16 & 5895 & 1.45 -- 3.71 & $-$4 -- $+$ 41 & 1\\
             &14 & 7543 & 0.29 -- 3.89 & $-$25 --$+$ 45 & 1\\
             & 3 & 5500 & 1.7  -- 3.1  & 18 -- 42       & 2\\
             & 2 & 6800 & 2.4  -- 3.1  & 22 -- 39       & 2\\      
%            & 8 & 4450 & 1.39 -- 2.41 & 18 -- 41       & Schulte-L 83\\
             & 3 & 5500 & 5.13 -- 5.47 & 35 -- 38       & 3\\
             & 3 & 6800 & 4.52 -- 4.98 & 33 -- 36       & 3\\
             & 3 & 5895 & 2.98 -- 3.13 & 17 -- 20       & 4\\
             & 1 & 6800 & 4.0          & 14             & 5\\
             & 7 & 6800 & 2.48 -- 4.03 & 17 -- 26       & 6\\
T Tau        & 8 & 5895 & 0.94 -- 1.53 & 95 -- 100      & 1\\
             & 7 & 7543 & 0.84 -- 1.50 & 94 -- 101      & 1\\
             & 3 & 5500 & 0.75 -- 1.2  & 90 --  98      & 2\\
             & 2 & 6800 & 0.65 -- 1.14 & 91 -- 95       & 2\\
             & 1 & 5895 & 0.75         & 100            & 4\\
%            & 1 & 4700 & 1.26         & 99             & M\'enard 92\\
SU Aur       & 4 & 5895 & 0.09 -- 0.21 & 106 -- 175     & 1\\
             & 3 & 7543 & 0.09 -- 0.35 & 101 -- 142     & 1\\
             & 2 & 5500 & 0.13 -- 0.25 & 119 -- 138     & 2\\
             & 1 & 6800 & 0.11         & 98             & 2\\  
%            & 7 & 4450 & 0.15 -- 0.36 &  76 -- 173     & 3\\
             & 3 & 5500 & 0.48 -- 0.66 & 127 -- 132     & 3\\
             & 3 & 6800 & 0.46 -- 0.58 & 119 -- 128     & 3\\
             & 1 & 5895 & 0.57         & 103            & 4\\
FU Ori       & 7 & 5895 & 0.68 -- 0.83 & 125 -- 135     & 1\\
             & 7 & 7543 & 0.68 -- 0.83 & 127 -- 133     & 1\\
CO Ori       & 1 & 5895 & 1.61         &  6.1           & 1\\
             & 1 & 7543 & 1.59         &  6.1           & 1\\
             & 1 & 5895 & 2.76         & 178            & 4\\
%            & 4 & 4700 & 1.48 -- 2.12 & 172 -- 175     & Menard 92\\
             & 8 & 6800 & 1.75 -- 2.60 & 156 -- 169     & 6\\
%DR Tau      &12 & 4450 & 0.12 -- 0.92 & 48 -- 167      & 3\\
DR Tau       & 1 & 5500 & 1.28         & 140            & 3\\
             & 1 & 6800 & 1.1          & 139            & 3\\
             & 3 & 5250 & 0.38 -- 0.59 & 124 -- 152     & 7\\ 
             & 4 & 6800 & 0.52 -- 0.81 & 125 -- 139     & 6\\
RW Aur       & 2 & 5895 & 0.68 -- 1.44 & 53 -- 94       & 1\\
             & 2 & 7543 & 0.68 -- 0.85 & 49 -- 104      & 1\\
             & 3 & 5500 & 0.6  -- 0.9  & 101 --103      & 2\\
             & 2 & 6800 & 0.8 -- 0.9   &  98 -- 102     & 2\\
%            &11 & 4450 & 0.18 -- 0.91 &  72 -- 126     & 3\\
             & 4 & 5500 & 0.45 -- 1.12 &  78 -- 105     & 3\\
             & 4 & 6800 & 0.41 -- 0.92 &  82 -- 100     & 3\\
             & 1 & 5895 & 0.79         &  77            & 4\\   
GW Ori       & 1 & 5895 & 0.21         & 102.6          & 1\\
             & 1 & 7543 & 0.35         & 102.0          & 1\\
             & 1 & 5500 & 0.52         & 131            & 2\\
             & 1 & 6800 & 0.5          & 116            & 2\\
             & 1 & 5895 & 2.64         & 151            & 4\\
%            & 4 & 4700 & 0.44 --1.01  & 92 -- 102      & Menard 92\\      
%UX Tau  AB  & 2 & 5895 & 0.62 -- 0.63 & 51 -- 64       & 1\\
%        AB  & 2 & 7543 & 0.56 -- 0.63 & 61 -- 67       & 1\\
UX Tau       & 1 & 5895 & 0.45         & 68             & 4\\
%       A    & 1 & 4700 & 0.69         & 67             & Menard 92\\
%       AB   & 1 & 4700 & 0.90         & 72             & Menard 92\\
BP Tau       & 2 & 5895 & 0.17 -- 0.36& 83.7            & 1\\
             & 2 & 7543 & 0.19 -- 0.29 & 69 -- 80       & 1\\
             & 1 & 5895 & 0.31         & 93             & 4\\
\hline
\end{tabular}
\\
\noindent
$^1$ Bastien (1982)
$^2$ Hough et al. (1981)
$^3$ Schulte-Ladbeck (1983)
$^4$ Bastien (1985)
$^5$ Bergner et al. (1987)
$^6$ Oudmaijer et al. (2001)
$^7$ Drissen et al. (1989)
%\end{minipage}
\end{table}

\begin{figure*}
\mbox{
\epsfxsize=0.33\textwidth\epsfbox{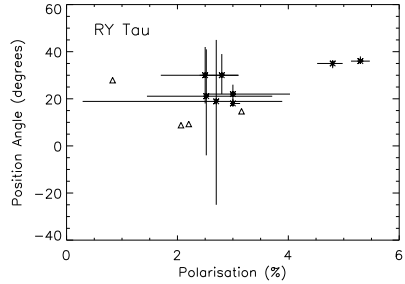}
}
\caption{Plot of the current (open triangles) and previous (crosses) continuum \% Pol and PA for 
RY Tau. Note that the large crosses do not represent error bars, but short-term 
polarisation variability with $N$ measurements, where $N$ is given in the 4th Column of Table~\ref{t_earlier}.
}
\label{f_indpolrytau}
\end{figure*}

\begin{figure*}
\mbox{
\epsfxsize=0.33\textwidth\epsfbox{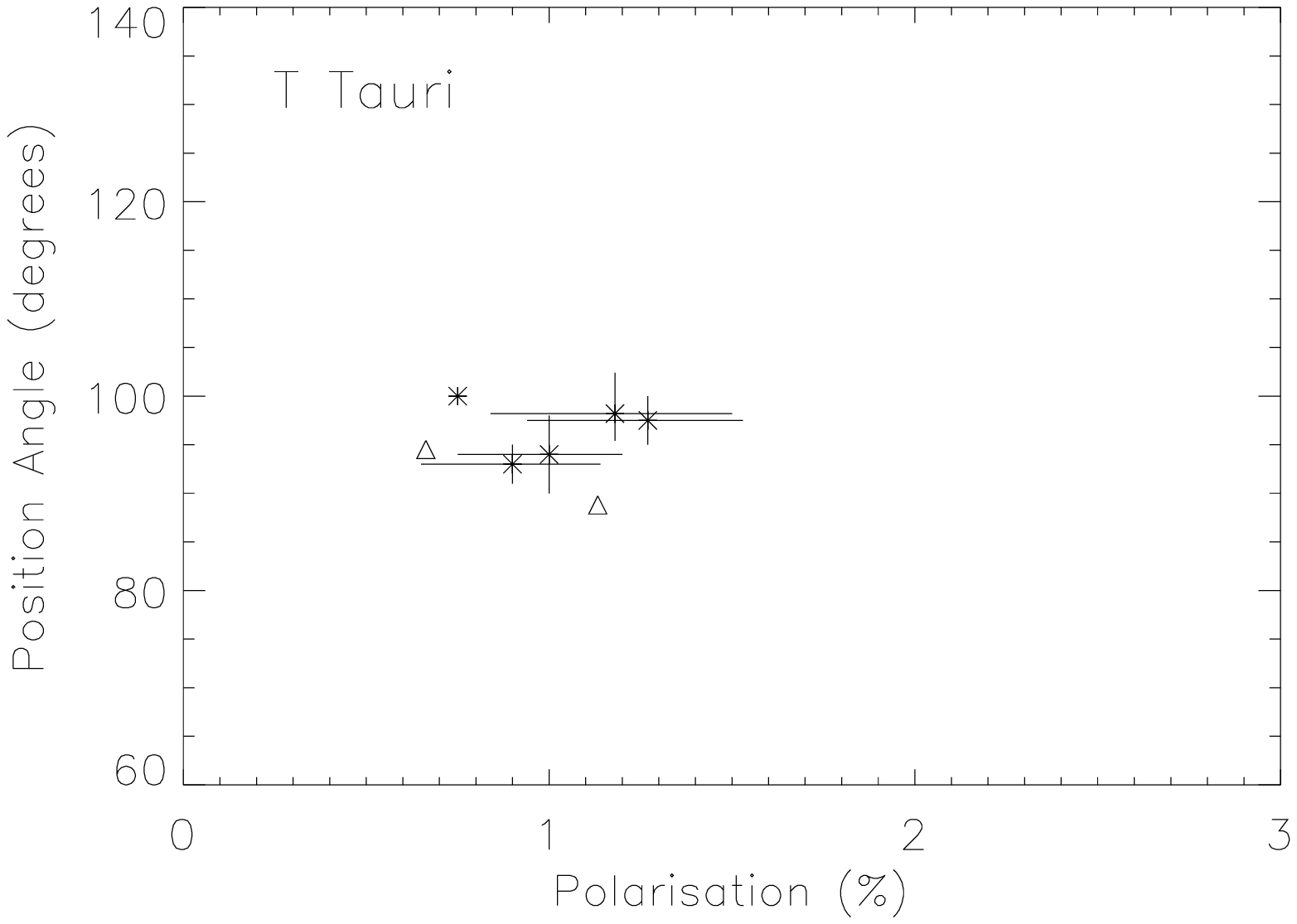}
\epsfxsize=0.33\textwidth\epsfbox{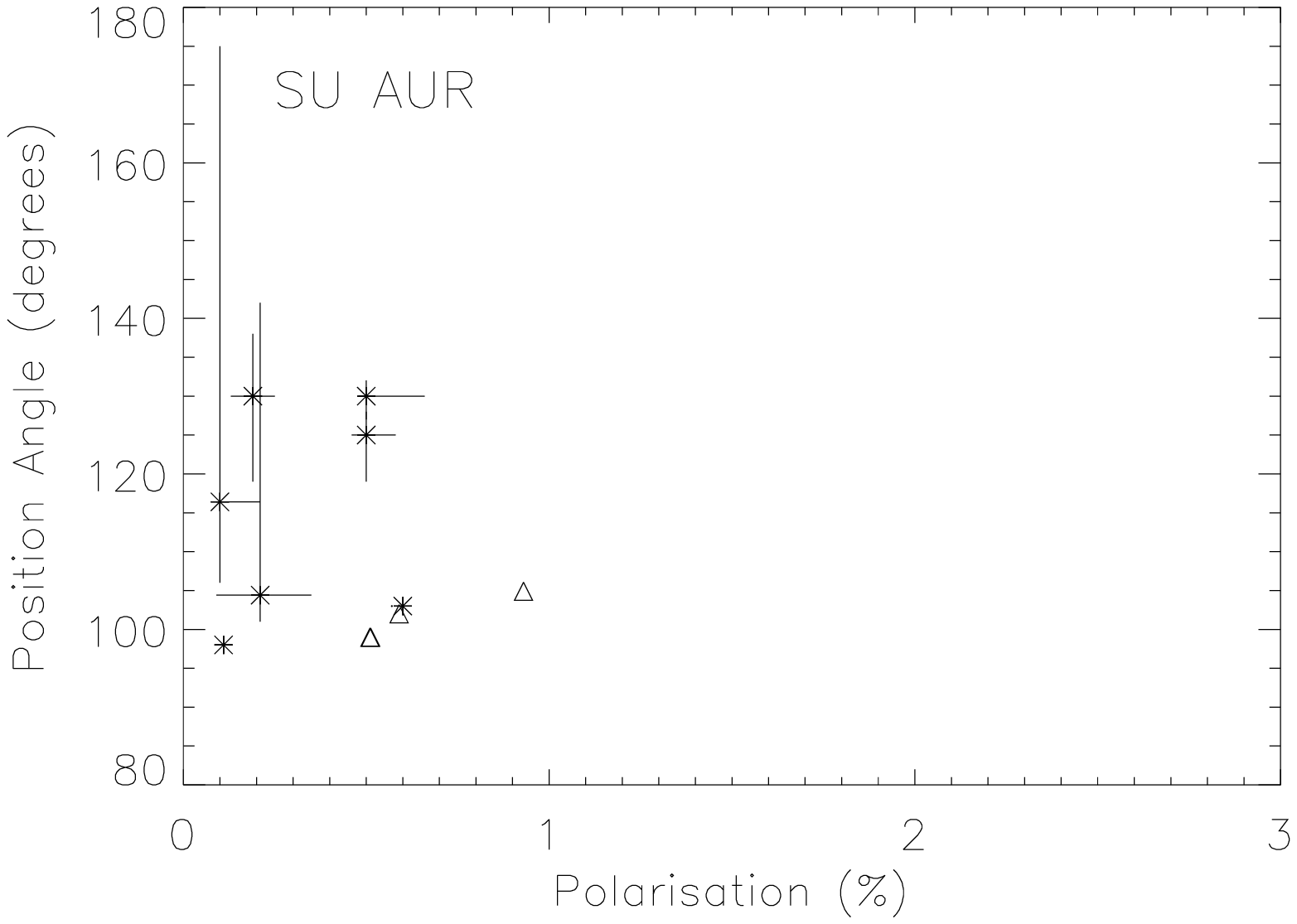}
\epsfxsize=0.33\textwidth\epsfbox{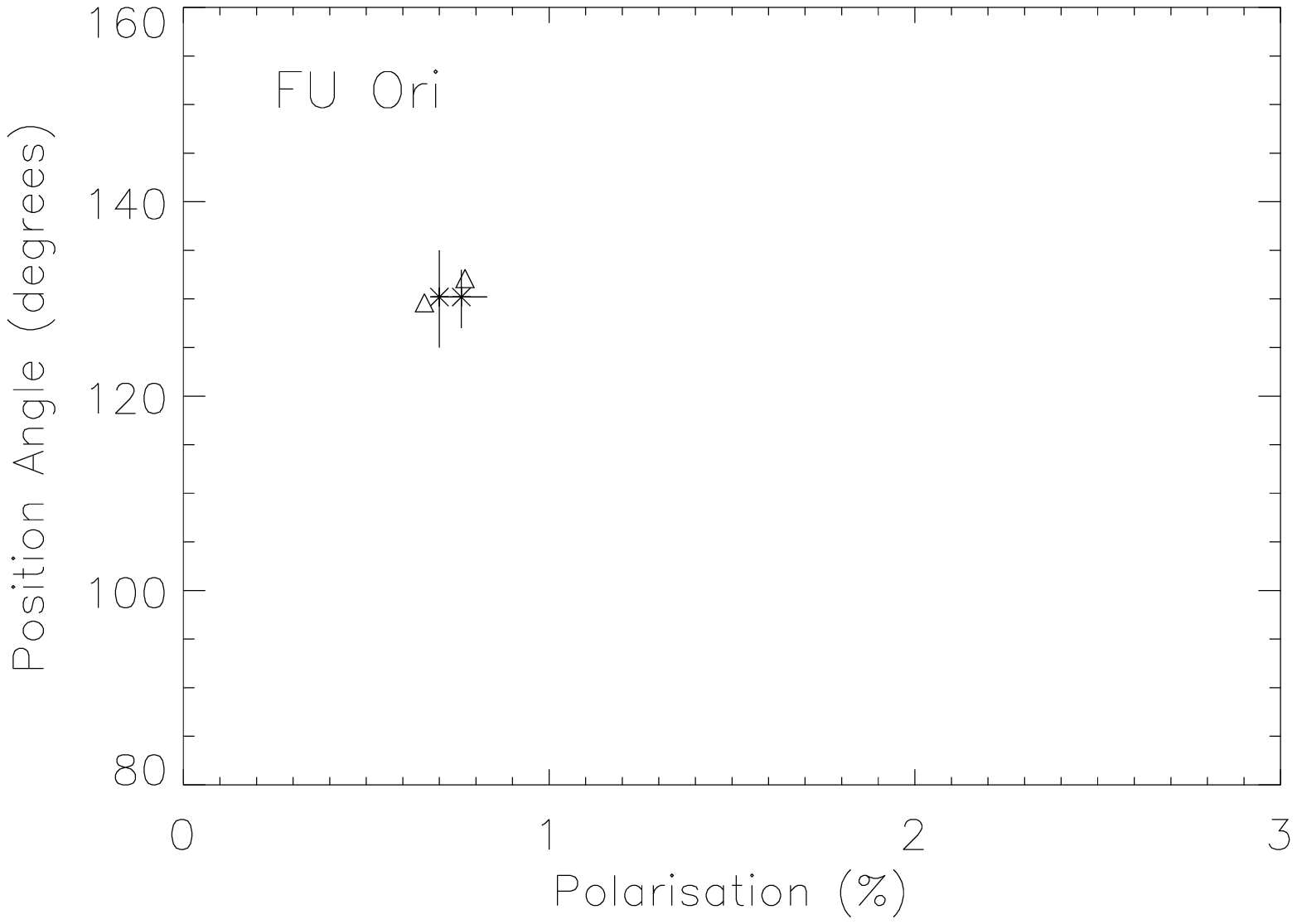}
}
\mbox{
\epsfxsize=0.33\textwidth\epsfbox{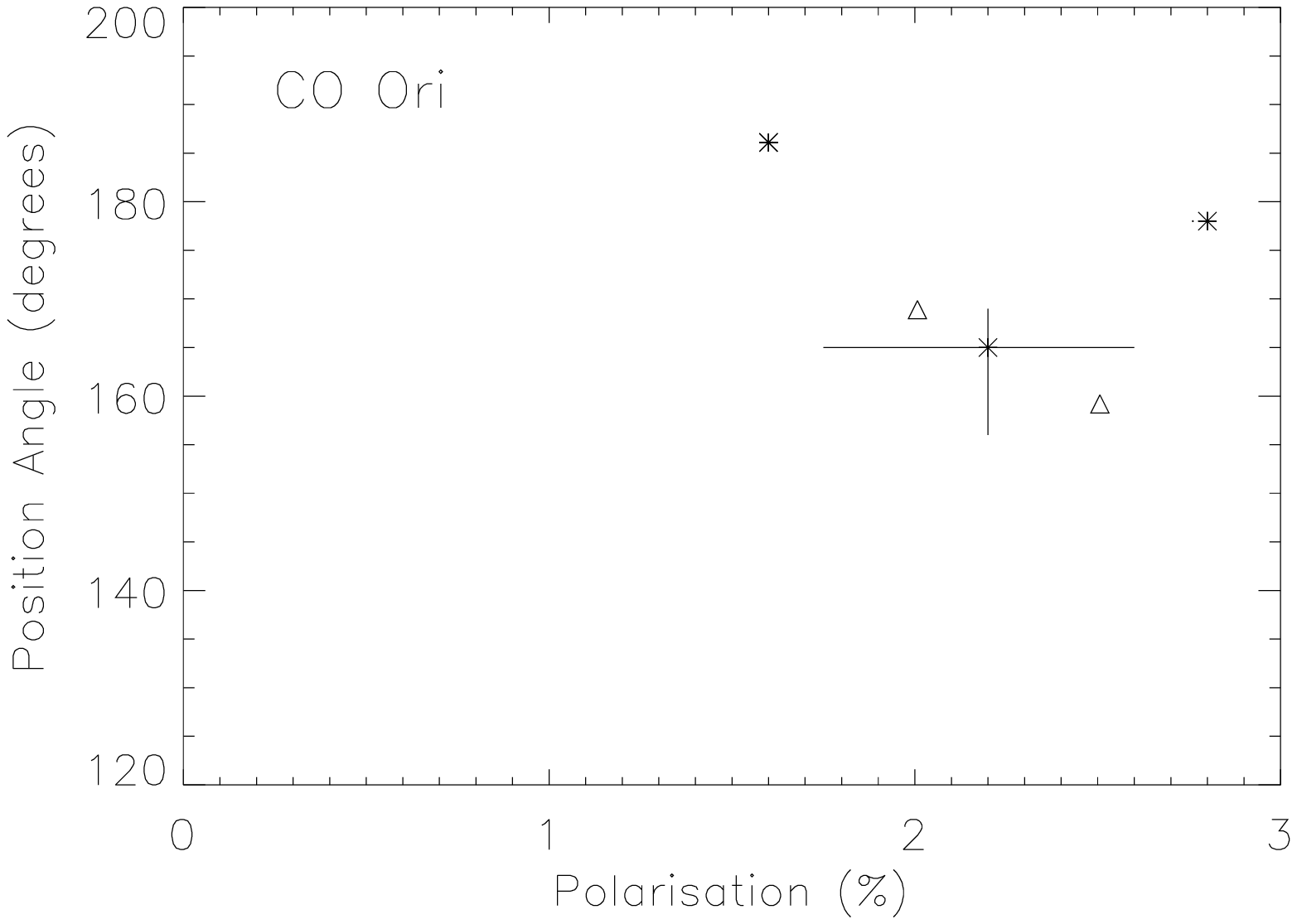}
\epsfxsize=0.33\textwidth\epsfbox{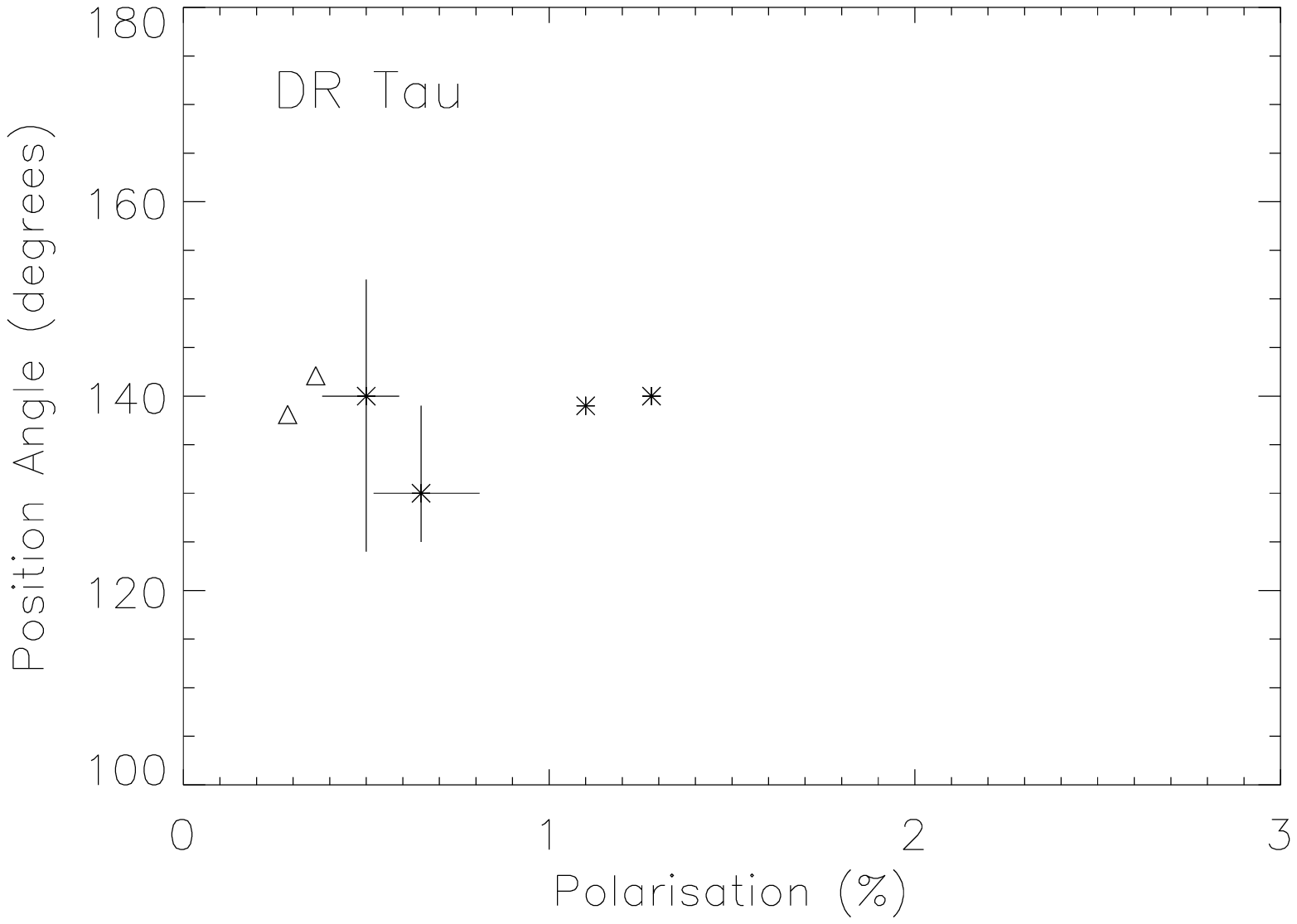}
\epsfxsize=0.33\textwidth\epsfbox{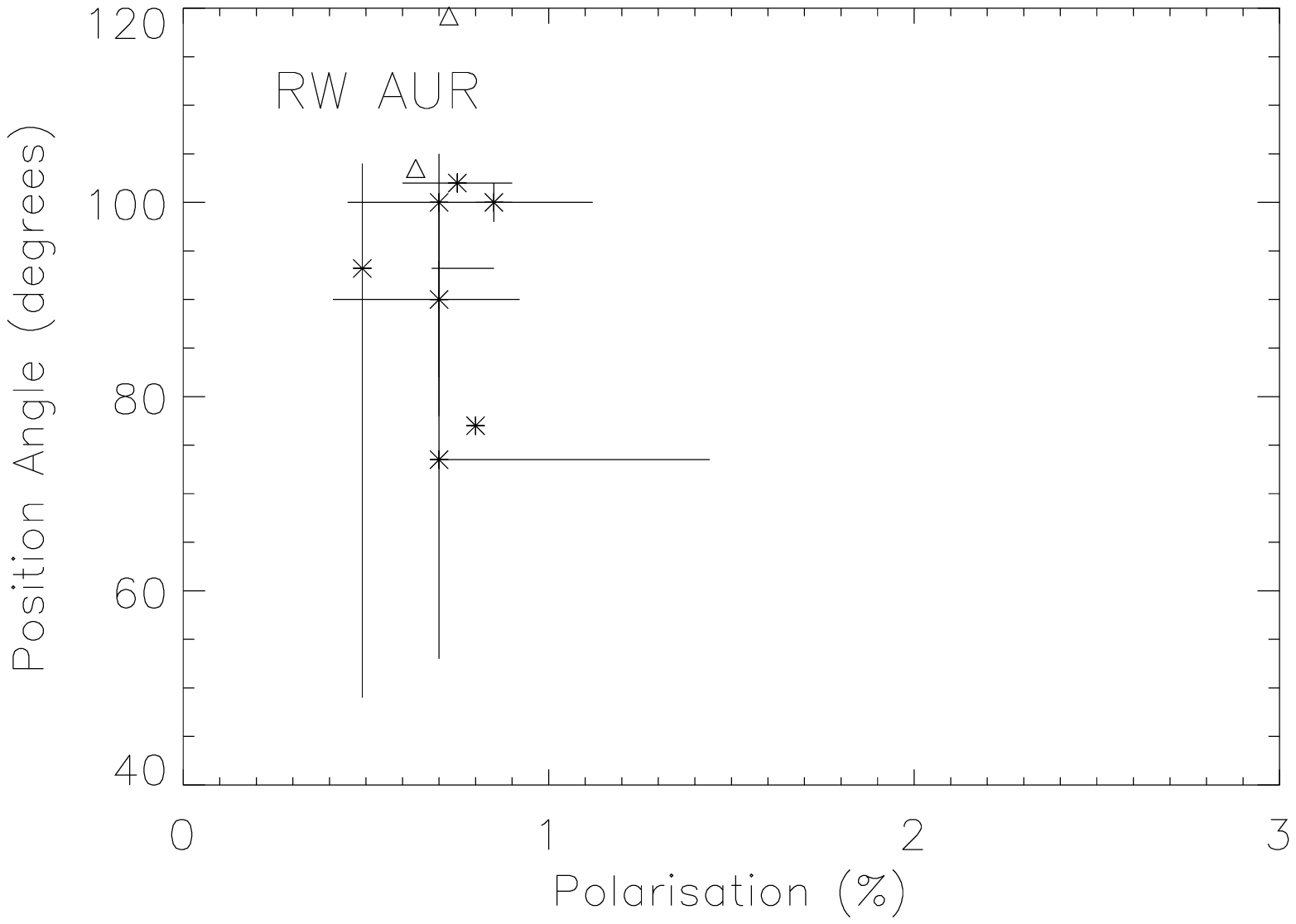}
}
\mbox{
\epsfxsize=0.33\textwidth\epsfbox{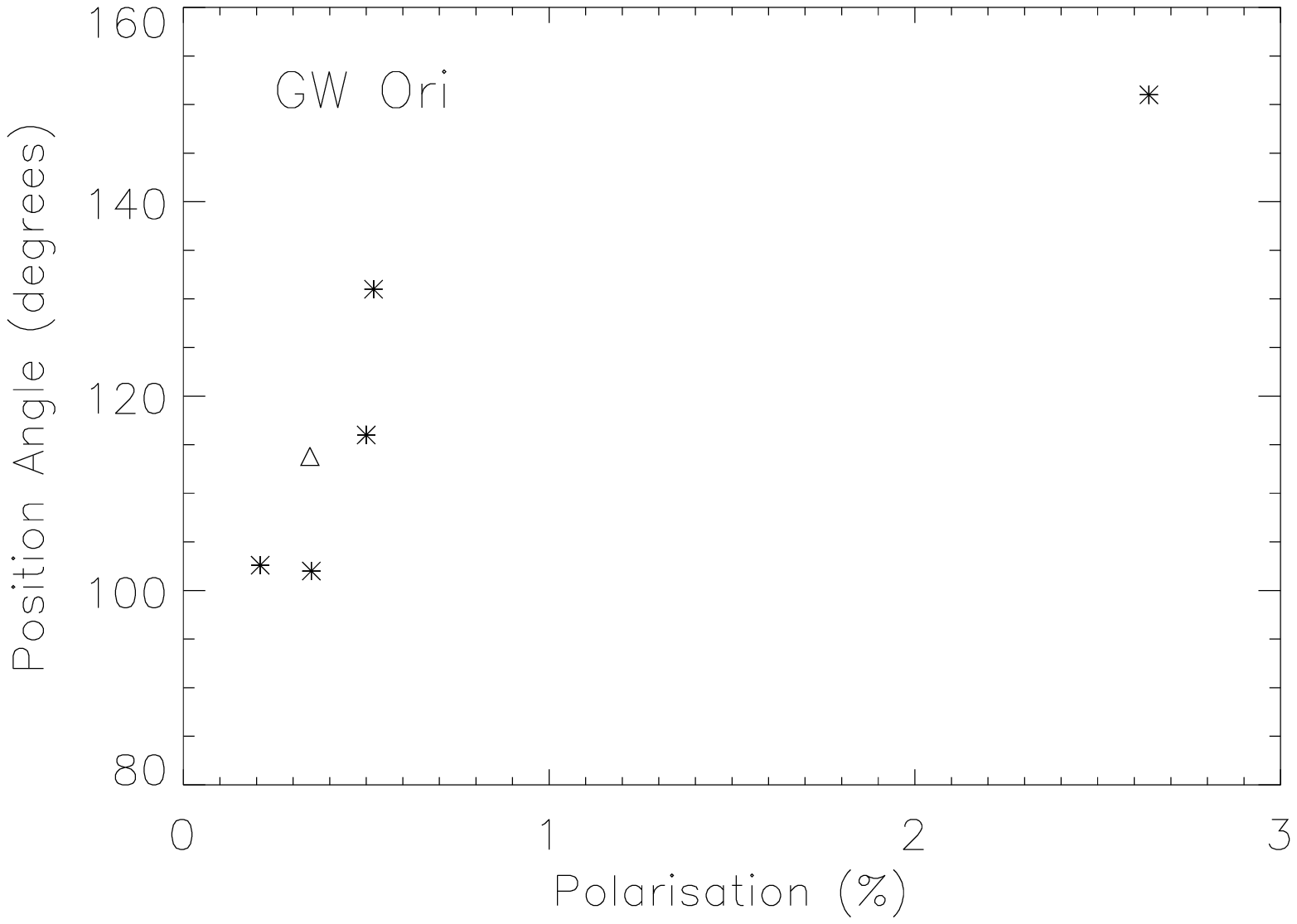}
\epsfxsize=0.33\textwidth\epsfbox{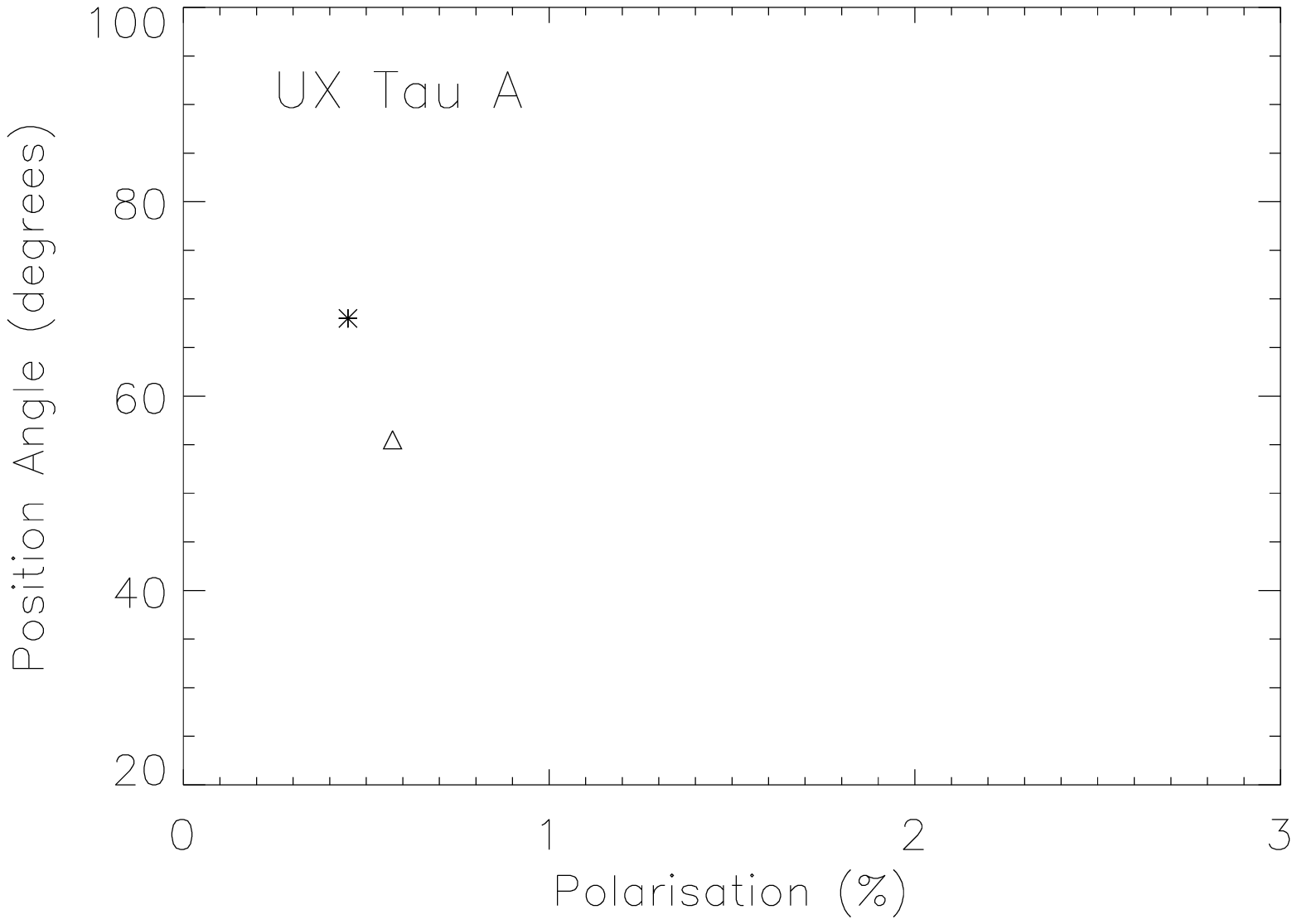}
\epsfxsize=0.33\textwidth\epsfbox{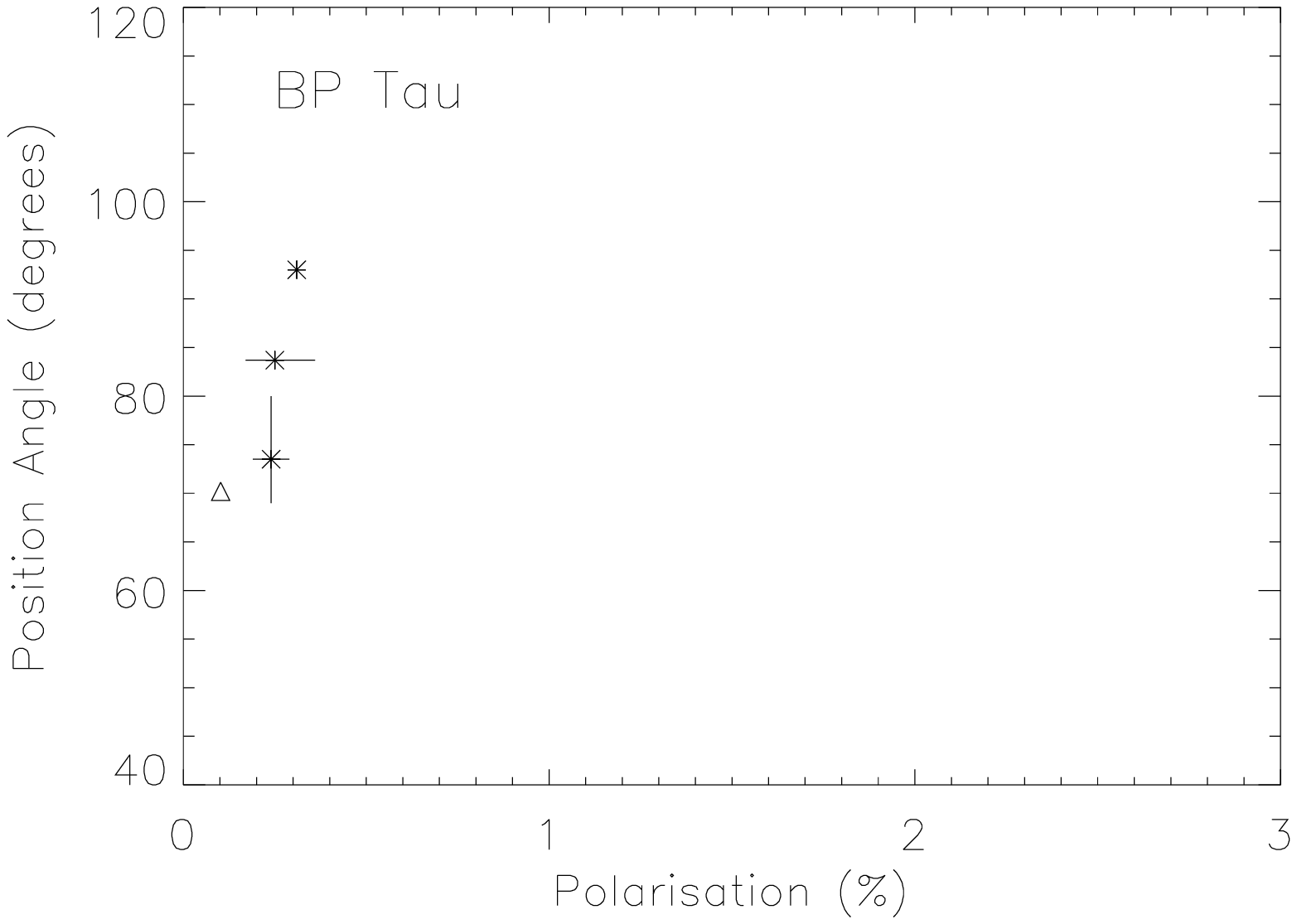}
}
\caption{Plots of the current (open triangles) and previous (stars) continuum polarisation and PA for 
the other T Tauri stars, with the y and x-axes fixed to respectively a PA range of 100\degree, and 
a \% Pol of 3\degree. Note that the large crosses do {\it not} represent error bars, but short-term 
polarisation variability over $N$ measurements, with $N$ given in the 4th Column of Table~\ref{t_earlier}.
}
\label{f_indpol}
\end{figure*}

\section{T~Tauri star results}
\label{s_ind}

We searched the literature for relevant information on individual objects.
Since this is the first extensive line polarimetry study on T~Tauri stars, the polarisation 
review consists of continuum measurements only. These are plotted 
in Fig.~\ref{f_indpol}, and summarised in Table~\ref{t_earlier}.
We have restricted our review to $V$ and $R$ bands only, and not considered wavelengths shorter than 5500 \AA, or 
longer than 7600 \AA. 

In Sect.~\ref{s_inner}, the polarisation information is complemented with information from imaging studies. 
We stress that for an approximately single-scattering disk, the PA is expected to 
lie orthogonal to the imaged disk. For a more opaque disk however, multiple scatterings 
in the vertical disk direction enforce the PA to align with the disk (Angel 1969; 
Bastien \& M\'enard 1988). 
The imaging studies comprise both millimetre (mm) studies, 
and near-infrared (NIR) imaging with a coronograph, or interferometry. The information we 
obtain is that of disk position angles and inclinations. The goal of the position angles 
is to compare them with the PAs we derive from the line polarimetry. From the 
inclinations, we can constrain disk inner hole sizes using the line polarimetry predictions 
of Vink et al. (2005). In Sect.~\ref{s_fore}, we change focus to the larger scale, associated with 
the orientation of the environment's magnetic field direction.

\subsection{The inner disk results}
\label{s_inner}

\subsubsection{RY~Tau}

RY~Tau has a relatively low level of veiling at optical wavelengths, i.e. $\le$ 0.1 
(e.g. Basri, Martin \& Bertout 1991), however the veiling at infrared wavelengths is markedly 
higher, i.e. $>$ 0.8 (Folha \& Emerson 1999). This mismatch in veiling is typical for T~Tauri stars, and 
it suggests the object may still be actively accreting, whilst exhibiting a relatively low level of optical veiling.
RY~Tau has a rotational velocity of $v \sin i = 55 \pm 3$ \kms\ (Mora et al. 2001), but the 
period of 5.6 days, as reported by Herbst et al. (1987), has never been confirmed.
From millimetre mapping, Koerner \& Sargent (1995) infer a disk PA of 48\degree\ $\pm$ 5\degree, 
with an inclination angle $i$ = 25\degree.
Kitamura et al. (2002) report a disk with a PA of 59\degree $\pm$ 7\degree\ and an inclination angle of $i$ = 27\degree\ $\pm$ 7\degree. 
They also derive 
a disk inner radius of $R_{\rm in}$ = 0.14 AU = 10 $\rstar$ from their modelling.
Finally, Akeson et al. (2003) perform NIR interferometry, and they find a disk at a 
PA of 62\degree $\pm$ 10\degree\ and an inclination of 
$i$ = 30\degree.

The linear percentage continuum polarisation has been reported to vary 
between 1 and 4\%, and its PA between $-$20 and 45\degree, with \%Pol 
$\simeq 3$ and PA $\simeq$ 20\degree\ being typical ``median'' values (see Table~\ref{t_earlier}).\\

{\it Our data:~}
We have measured a continuum polarisation between about 0.8 and 3.1\% at a somewhat variable PA of approximately 
20\degree (Fig.~3). This is in keeping with earlier continuum measurements.
We have already reported on the line polarimetry of RY~Tau in Vink et al. (2003):  
the line polarisation data indicate the presence of a rotating disk, and we argued for 
an intrinsic PA of 147\degree\ or 163\degree. 
The expected single scattering disk PAs are perpendicular to this at 57\degree\ and 73\degree, respectively. 
These angles are comparable to those derived from imaging (PA = 62\degree).

\subsubsection{T~Tau}

T Tau is a triple system (Dyck, Simon \& Zuckerman 1982; Ghez et al. 1991; HB; Koresko 2000), and T~Tau North is  
the optically visible prototype T~Tauri star. Smirnov et al. (2003) have possibly detected a small magnetic field.
The star's rotational velocity is $v$sin$i$ = 19.5 $\pm$ 2.5 \kms (Bertout et al. 1986), and the rotational period is 
2.80 days (Herbst et al. 1987).
Hogerheijde et al. (1997) and Akeson, Koerner \& Jensen (1998) detect a disk from mm observations and conclude that the disk 
is observed nearly face-on, i.e. at a low inclination. 
Johns \& Basri (1995) had already claimed a low inclination for the system on the 
basis of the \ha\ correlation matrix, which was confirmed by Calvet et al. (1994) who 
modeled the spectral energy distribution (SED). Most recently, Akeson et al. (2002) have 
performed IR interferometry, confirming the low inclination of the system:
$i$ = 29\degree$_{-15}^{+10}$ at a PA = 132\degree$_{-20}^{+13}$. \\

{\it Our data:~} 
The line polarisation data for this star (Fig.~\ref{f_noline}) do show variability. 
The data from 2001 may show a weak depolarisation, but those from 2003 do not.
This change may simply be due to a fall in the amount of ionised gas close to the  
star that can scatter and polarise the continuum photons and emit in \ha\ 
(the \ha\ equivalent width decreases too, cf. the discussion on the Herbig Be star MWC~166 in Vink et al. 2002). 
In any case, the absence or weakness of the depolarising line effect 
is not surprising given T Tau's pole-on orientation, as circular symmetry on the sky results 
in a cancellation of the polarisation vectors.

\subsubsection{SU~Aur}

The relatively massive T~Tauri star SU~Aur has an uncertain period of in between  
1.7 (DeWarf et al. 2003) and 2.7 days (Herbst et al. 1987; Unruh et al. 2004). 
The rotational velocity is $v$sin$i$ = 59 $\pm$ 1 \kms.
Akeson et al. (2002) image a disk at a PA of 127\degree$_{-9}^{+8}$. They derive an  
inclination of $i$ =  62\degree$_{-8}^{+4}$, which is consistent with the Doppler imaging 
studies of Petrov et al. (2001) and Unruh et al. (2004) which also favour $i$ 
between 50\degree\ and 70\degree. Akeson et al. (2002) also model the disk inner rim and 
find a hole of radius 0.05 -- 0.08 AU, which corresponds to $\simeq$ 2.7 -- 4.3 stellar 
radii. This compares well with the expected corotation radius of 2.5 -- 3 stellar radii. \\

{\it Our data:~}
We have measured a continuum polarisation between about 0.5 and 0.9\% at a more or less 
fixed PA of $\simeq$ 100\degree. 
Although the PA is variable, the 
100\degree\ seen here appears to be common for this object, since it generally experiences
a PA in the range 100\degree\ -- 130\degree. 

The single $QU$ loop, at a preference PA of $\simeq$ 120\degree, may highlight two issues. 
First, the PA of the intrinsic polarisation aligns with the PA found from mm imaging, which would
suggest the presence of an optically thick multi-scattering envelope, such that the PA of the polarisation aligns 
with the disk PA. 

Second, the sheer presence of a {\it single} loop suggests the presence of a disk 
inner hole. 
This deduction is based on the Monte Carlo models by Vink et al. (2005) which show that there is a 
marked difference in polarisation changes across line profiles of 
line emission originating from a point source -- resembling a situation where the photons are 
scattered off a disk with a large inner hole -- and a finite-sized star -- where the inner hole is small. 
In between these two extremes, there is transitional behaviour which can also be used to 
constrain the size of the disk inner hole, provided the disk inclination is known.
Since SU~Aur is a system with a well-known inclination, $i$ $\simeq$ 60\degree,  we 
can use the predictions by Vink et al (2005) to estimate a lower limit to the disk inner radius: 
$\Rin$ $>$ 3 $\Rstar$. 
This is consistent with the inner radius found by Akeson et al. (2002).

\subsubsection{FU Ori}
\label{s_fu}

FU Ori is the prototype of the outbursting low mass young stellar objects. 
These stars show large increases in optical brightness of $\simeq$ 5 magnitudes on timescales
ranging from less than a year, to more than a decade (e.g. Herbig 1977).
The frequency of these outbursts is uncertain, but it is believed that these outbursts 
may represent the principal mode of accretion onto young stars (e.g. Hartmann \& Kenyon 1985).
The deeply seated P~Cygni absorption is usually interpreted as being due 
to an almost pole-on system. 
Recently, Wang et al. (2004) have detected a faint red star in the apparent vicinity 
of FU~Ori. This is possibly a close companion at a PA of 161\degree\ $\pm$ 3\degree\ with 
a separation of about 225 AU.  

Malbet et al. (1998) resolved FU Ori, but could not unravel its morphological structure. They also  
modelled the SED, and found an inclination of 30\degree. 
More recently, Malbet et al. (2005) have performed 2.2 $\micron$ interferometry, and 
they now find actual physical parameters for the disk PA and inclination, using two different models. 
In the first model (disk only), the PA is 47\degree\ $\pm$ 10\degree\ and $i$ = 55\degree $\pm$ 7\degree. 
In the second model (a disk and a bright spot), they find a PA 
of 8\degree\ $\pm$ 21\degree and $i$ = 46\degree $\pm$ 10\degree. \\

{\it Our data:~}
We have measured a continuum polarisation of about 0.7\% at a PA of approximately 130\degree. 
It appears that the level of 
\%Pol as well as the PA remain constant. 
Despite the low S/N, the line polarimetry shows a clear occurrence of the ``McLean effect'' 
(McLean 1979), where blue-shifted line absorption (in Stokes $I$) is accompanied 
by a change in the polarisation percentage and/or PA over its width. 
This change in polarisation/PA corresponds to an excursion 
in the $QU$ plane at $\simeq$ 45 degrees, which defines the intrinsic plane of the continuum polarisation
(independent of foreground polarisation).
The reasoning is that the blue-shifted absorption mostly removes direct stellar continuum light, which
is unscattered and unpolarised. As a result, the (few) blue-shifted photons that do reach the observer 
are selectively more polarised, defining the intrinsic plane of continuum polarisation.   
Since the disk of FU~Ori is optically thick, with a high accretion rate (of up to 10$^{-4}$ \msun\ yr$^{-1}$), we would 
expect the PA of the optically thick 
disk on the sky to align with the PA from the line excursion. It appears that the PA we derive from the $QU$ plane
is consistent with the PA of 47\degree\ $\pm$ 10\degree\ from the disk model by Malbet et al (2005).

\subsubsection{CO Ori}

Despite its relative brightness, CO~Ori appears to be poorly studied. 
Reipurth et al. (1996) note that the pronounced P~Cygni profile of CO~Ori (not seen in our data) 
is reminiscent of FU Ori, as well as its relatively early spectral type. \\

{\it Our data:~}
We have measured a continuum polarisation of about 2.0\% at a PA of approximately 160\degree, which 
is consistent with earlier continuum measurements (Table~\ref{t_earlier}, Fig.~4). 
Despite the relatively low S/N, we tentatively infer an intrinsic PA of $\simeq$ 60\degree\ from the excursion 
in the $QU$ plane. When the larger scale disk of this object will be imaged, we expect the imaged disk 
to have a PA of $\simeq$ 60\degree, if it is an optically thick, multiple scattering disk.

\subsubsection{DR Tau}
\label{s_dr}

DR~Tau exhibits strongly variable emission lines (e.g. Smith et al. 1999) which may signal both inflow and outflow. 
The Balmer lines are strongly asymmetric, peaking in the red, and do not resemble magnetospheric infall model predictions.
Alencar, Johns-Krull \& Basri (2001) propose the system is observed nearly pole-on, but 
Muzerolle, Calvet \& Hartmann (2001) suggest a high inclination ($i =$ 70\degree), and advocate a 
scenario in which most of the \ha\ is formed in the stellar wind, rather than in the 
accretion flow. The mm imaging data of Kitamura et al. (2002) seem to support the more edge-on viewing 
angle: $i$ $\simeq$ 67\degree\ at a PA of $\simeq$ 128\degree. They estimate a disk radius of 
about 200 AU, and they derive an inner radius of 10 $\Rstar$ based on disk models.
Kravtsova \& Lamzin (2002) argue that $i$ $<$ 60\degree, and 
Greaves (2004) models the HCO$^{+}$ lines, and she finds an inclination between 4 -- 9\degree. These results
appear to be in contradiction to the mm study by Kitamura et al. 
However, based on all other evidence, such as the low rotational velocity, $v$ sin$i$ $<$ 10 \kms, 
the period of 5.1 days (Johns \& Basri 1995) and the stellar size, we consider
the more pole-on orientation to be more plausible.

{\it Our data:~}
We have measured a continuum polarisation of about 0.3\% at a PA of approximately 140\degree (consistent 
with earlier continuum measurements, cf. Table~\ref{t_earlier}).
 
The line polarimetry data show the occurrence of the ``McLean effect'', where blueshifted line 
absorption (in Stokes $I$) is accompanied 
by a change in the \%Pol/PA over the same wavelength region as the P~Cygni 
absorption (cf. the case of FU~Ori). 
From the loop in the $QU$ plane, we infer an intrinsic PA of $\simeq$ 120 
degrees for the system, which is in good agreement with the PA of 128\degree\ found by Kitamura et al.  

\subsubsection{RW Aur A}

RW Aur is a resolved triple system (Ghez, Neugebauer \& Matthews 1993) with RW Aur A dominating the optical light. 
The star has a rotational velocity of $v$ sin$i$ = 19.5 $\pm$ 4.6 \kms (Hartmann et al. 1986), while  
Petrov et al. (2001) report a photometric period of in between 2.6 and 2.9 days. L\'opez-Mart\'in, Cabrit \& 
Dougados (2003)
infer a jet inclination angle of $\simeq$ 45\degree. 
From HST spectroscopy G\'omez de Castro \& Verdugo (2003) infer an inner disk radius of between 
2.7 $\rstar$ and the corotation radius of 6.1 $\Rstar$. \\

{\it Our data:~}
We have measured a continuum polarisation of about 0.7\% at a PA of approximately 110\degree. 

The line polarimetry data seem to show the occurrence of the ``McLean effect'', 
here not across a bona-fide P~Cygni profile, but nonetheless across envelope absorption.
From the loop in the $QU$ plane, we infer an intrinsic PA of $\simeq$ 115\degree\ for the system.  
When the larger scale disk of this object will be imaged, we expect the imaged disk 
to have a PA of $\simeq$ 115\degree, if it is an optically thick, multiple scattering disk.

\subsubsection{GW Ori}

GW Ori is a single-lined spectroscopic binary (Murdin \& Penston 1977) with a period of 
242 days, in a nearly circular orbit. The system is surrounded by a massive circumbinary disk. 
Most of the sub millimetre light originates at 500 AU (Mathieu et al. 1995).

The star has a rotational velocity $v$ sin$i$ of 40 -- 43 \kms (Bouvier et al. 1986; Hartmann et al. 1986), 
and a period of 3.2 days (Bouvier \& Bertout 1989), although this period has not been 
confirmed by Gahm et al. (1993). Bouvier \& Bertout (1989) infer an almost pole-on orientation
angle $i$ of 15\degree, while Mathieu, Adams \& Latham (1991) prefer $i$ $\simeq$ 27\degree, since only 
26 \lsun\ (out of a 70 \lsun\ total) is believed to be stellar, with the remainder of the system's 
luminosity attributed to accretion. 
From SED modeling, Mathieu et al. (1991) derive a gap in the disk over the range of 0.17 -- 3.3 AU, in broad 
consistency with the models of Artymowicz \& Lubow (1994) who predict a disk gap between 0.45 and 2.0 AU, 
due to the secondary star at a distance of 1.1 AU from the primary. Najita, Carr \& Mathieu (2003) derive a CO emission 
ring between an inner radius $\Rin$ of 0.07 AU (i.e. 3 $\Rstar$) and an outer radius of 0.6 AU, adopting $i$ $\simeq$ 
27\degree.

We note that the object appears to show (quasi)-algol type eclipses, which Shevchenko et al. (1998) attributed to a high 
binary inclination, with Lamzin et al. (1998) reporting $i$ $\simeq$ 83\degree.
For the PA of the object, Mathieu et al. (1995) report a value 
56\degree, but warn that the object is not well-resolved. \\

{\it Our data:~}
We have measured a continuum polarisation of about 0.3\% at a PA of approximately 114\degree.  
The line polarimetry data show a loop in the $QU$ plane, possibly at an intrinsic PA of approximately 60\degree, 
which may, or may not, be consistent with the value of 56\degree\ in Mathieu et al. (1995).

The gradual slope in the behaviour of 
the PA is reminiscent of the transitional PA behaviour between a finite-sized star and a point-source 
as predicted by Vink et al. (2005). If we assume the low inclination of 15\degree, 
and look up the transitional behaviour of the PA for this inclination in Vink et al. (2005), we find a 
disk inner hole size of $\simeq$ 4 $\Rstar$. For $i$ $\simeq$ 27\degree, the disk inner hole would be 
smaller, e.g. $\simeq$ 3 $\rstar$, consistent with the study by Najita et al. (2003). 
We point out that we have assumed that the line photons originate from the primary, and that the secondary is not contributing 
significantly to the line emission. 

\subsubsection{UX Tau A}

UX Tau A has a rotational velocity of $v$ sin$i$ $\simeq$ 26 \kms (Hartmann et al. 1986; Bouvier \& Bertout 1989), and a 
period of 2.7 days according to Bouvier \& Bertout 1989 (based 
on data from Mundt et al. 1983). Bouvier \& Bertout (1989) also report an inclination of 50 $\pm$ 6\degree\ for 
UX Tau A, but this number is most uncertain, given the uncertain period and stellar size 
(due to the uncertain luminosity). 
We have searched the literature for imaging data, but these seem to be unavailable at the present time. \\

{\it Our data:~}
We have measured a continuum polarisation of about 0.5\% at a PA of approximately 
65\degree\ (in agreement with earlier measurements, cf. Table~\ref{t_earlier}). 
We can infer an intrinsic PA of $\simeq$ 140\degree\ from the excursion in the $QU$ plane.
When the larger scale disk of this object will be imaged, we expect the imaged disk 
to have a PA of $\simeq$ 140\degree, if it is an optically thick, multiple scattering disk.

\subsubsection{BP Tau}

BP Tau was the first T~Tauri star found to have a large surface
magnetic field (of 2.6 kGauss, Johns-Krull et al. 1999). 
More recently, Symington et al. (2005) find a peak longitudinal field of 4 kGauss.
The star has a low rotational velocity, i.e. $v$ sin$i$ $<$ 10 \kms (Hartmann et al. 1986),
but reliable periods have not yet been reported. Sub-millimetre imaging in the CO line 
by Simon, Dutrey \& Guilloteau (2000) reveals a PA of 152\degree\ $\pm$ 3\degree, and an inclination 
$i$ = 30\degree$_{-2}^{+4}$. Dutrey, Guilloteau \& Simon (2003) confirm the low inclination, $i$ = 28\degree\
$\pm$ 2\degree, but report a PA = 57\degree\ $\pm$ 4\degree. \\

{\it Our data:~}
We have measured a small continuum polarisation of about 0.1\% at a PA of approximately 70\degree. 

The continuum in the $QU$ plane is not well-defined, due to the low S/N, but the data are good 
enough to show the presence of a loop in the $QU$ plane. 
Despite the relative faintness of the object, the loop in the $QU$ plane and the presence of a 
strong magnetic field, make BP~Tau an ideal target for a dedicated spectropolarimetry monitoring 
program for detailed studies of the magnetically controlled accretion 
scenario in pre-main sequence stars.

\subsection{The connection with the orientation of the global magnetic field}
\label{s_fore}

\subsubsection{The foreground polarisation of FU~Ori}

For FU~Ori, we have measured an intrinsic PA of 45\degree\ (Sect.~\ref{s_fu}), which 
is exactly perpendicular to the measured continuum PA at $\simeq$ 130\degree, 
which is constant over time, cf. Fig.~\ref{f_indpol}. 
This 90-degree offset makes it highly likely that there is a significant amount of ``crossed'' 
polarisation, i.e. the intrinsic and foreground PA are 90 degrees rotated from one 
another. The reason being that there is substantial extinction towards FU~Ori, which 
makes it unlikely that the foreground polarisation is low. 
If it is not low, it must have a PA that is 90-degrees rotated from the intrinsic PA, with a level of 
polarisation larger than that of the intrinsic polarisation. If this were not so, the measured PA could not 
be perpendicular to the intrinsic one. To be more precise: 
given the \av\ of 1.38 -- 2.66 to FU Ori, and 
using the standard relation between \av\ and $\%P$, $\frac{P^{\rm max}}{A_{\rm V}}$ = 3, due to 
Serkowski, Mathewson \& Ford (1975), the expected amount of foreground polarisation cannot exceed
$\simeq$ 8\%. Since the measured polarisation is only of the order of $\simeq$ 1\%, it suggests 
that the intrinsic $P$ could be up to $\simeq$ 7\%. 
We note that these values are maximal and uncertain (due 
to the uncertainties in the used Serkowski relation), but the inferred 
90-degree offset in the PA of the intrinsic and foreground polarisation, in addition to the 
significant \av, strongly suggests that the value of the intrinsic polarisation greatly 
exceeds the measured one. We anticipate that this crossed polarisation is not a coincidence, but that it is related 
to the star formation process itself (see later).

\subsubsection{The foreground polarisation of DR~Tau}

For DR~Tau, we have measured an intrinsic PA of 120\degree (Sect.~\ref{s_dr}), which is almost 
parallel to the measured continuum PA at $\simeq$ 130\degree -- 140 \degree, cf. 
Fig.~\ref{f_indpol}. Although the intrinsic and measured PA are parallel for DR~Tau, whilst they were 
perpendicular for FU~Ori, the reasoning is the same. $QU$ vector addition of intrinsic and foreground 
polarisation can either result in a measured polarisation seen in a particular quadrant of the $QU$ plane, or 
on the opposite quadrant.
The consistency between the intrinsic and measured PA could suggest a very low foreground contribution, 
or, given the \av\ of 1.0, more likely to be crossed polarisation -- similar to the case of FU Ori. 
%Given the \av\ of 1.0, we consider the latter option more likely. 
Using the standard relation between \av\ and polarisation, one would
expect up to 3\% foreground polarisation.
Since the foreground PA is inferred to be at $\simeq$ 40\degree, whilst the PA of the 
intrinsic and measured polarisations are at $\simeq$ 130\degree, the intrinsic polarisation may  
exceed the amount of foreground polarisation by some amount. We argue that 
this cross-foreground effect for DR~Tau is probably as real as for the case outlined above, although the evidence for it 
is not as strong as for FU~Ori, for which its \av, and hence its expected polarisation is much larger.

\begin{figure*}
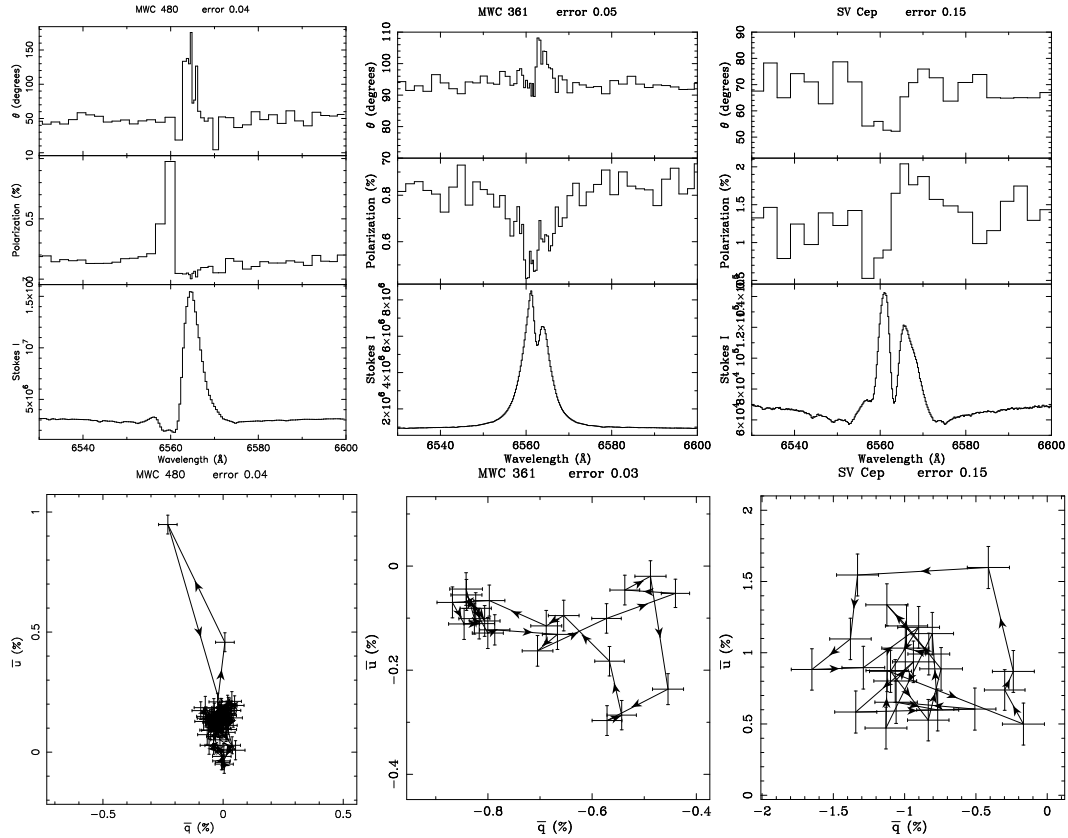

\mbox{
\epsfxsize=0.26\textwidth\epsfbox{m480.ps}
\epsfxsize=0.26\textwidth\epsfbox{m361.ps}
\epsfxsize=0.26\textwidth\epsfbox{sv.ps}
}
\mbox{
\epsfxsize=0.26\textwidth\epsfbox{m480_qu.ps}
\epsfxsize=0.26\textwidth\epsfbox{qum361_03.ps}
\epsfxsize=0.26\textwidth\epsfbox{qu_sv.ps}
}
\caption{Triplots and $QU$ plots of 3 Herbig Ae/Be stars that show a line effect.}
\label{f_yesherbig}
\end{figure*}

\begin{figure*}
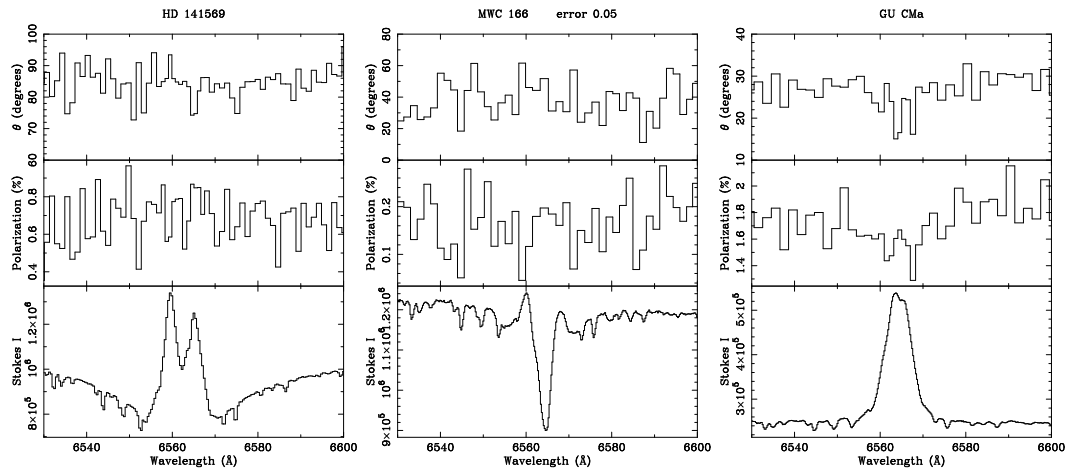

\mbox{
\epsfxsize=0.26\textwidth\epsfbox{hd141569.ps}
\epsfxsize=0.26\textwidth\epsfbox{m166_1.ps}
\epsfxsize=0.26\textwidth\epsfbox{gu1.ps}
}
\caption{Triplots and $QU$ diagrams of a few Herbig Ae/Be stars without the detection of a line effect.}
\label{f_noherbig}
\end{figure*}

\section[]{The Herbig Ae/Be stars}
\label{s_herbig}

The data of the complementary Herbig Ae/Be stars are presented in Figs.~\ref{f_yesherbig} and~\ref{f_noherbig}, 
as triplots and loci in the $QU$ plane, for objects that respectively do and do not show line effects.
Note that these Herbig Ae/Be star data only represent a fraction of our total Herbig Ae/Be star database
that was published in Vink et al. (2002). 

\begin{table*}
\begin{minipage}{\linewidth}
\renewcommand{\thefootnote}{\thempfootnote}
\caption{Herbig Ae/Be Targets. 
The integration times (column 6) denote the total exposures. 
The continuum PA and its error are indicated in column (8), whilst column (9) indicates whether we have 
good \ha\ line polarisation data; the $+$ indicates there is a plot of the  epoch.
Finally, column (10) indicates a measure of the sky PA derived from line excursions. 
}
\label{t_conth}
\begin{tabular}{lclcclcrlc}
\hline
Name & {\it V} & Spec. Tp & Type of object & Date & Exposure(s) & $P_{\rm cont}^{\rm R}$ (\%) & $\Theta_{\rm cont}^{\rm R}$ (\degree) & \ha\ data? & $\Theta_{\rm intr}^{\rm R}$ \\
\smallskip\\
 (1)   &  (2)  &  (3) & (4)  &  (5)  & (6)   & (7)  & (8)   & (9)  & (10) \\
\hline
MWC 480     & 7.7 & A5   & HAe  &26-12-01& 16$\times$75       & 0.176 $\pm$ 0.005 & 63.0 $\pm$ 0.8 & yes & 55\degree \\
            &     &      &      &26-12-01& 16$\times$30       & 0.173 $\pm$ 0.011 & 60.2 $\pm$ 1.8 & yes & \\
            &     &      &      &27-12-01& 16$\times$45       & 0.185 $\pm$ 0.006 & 59.7 $\pm$ 1.0 & yes & \\
            &     &      &      &27-12-01& 8$\times$45,8$\times$90   & 0.226 $\pm$ 0.017 & 56.9 $\pm$ 2.2 & yes & \\
            &     &      &      &28-12-01& 20$\times$60,8$\times$90  & 0.281 $\pm$ 0.009 & 61.8 $\pm$ 0.9 & yes & \\
            &     &      &      &28-12-01& 8$\times$45,16$\times$60  & 0.297 $\pm$ 0.010 & 64.8 $\pm$ 1.0 & yes & \\
AB Aur      & 7.1 & A0   & HAe  &27-12-01& 20$\times$30       & 0.109 $\pm$ 0.007 & 55.3 $\pm$ 1.7 & yes & \\
            &     &      &      &27-12-01& 16$\times$30       & 0.102 $\pm$ 0.008 & 57.0 $\pm$ 2.2 & yes & 160\degree\\
            &     &      &      &27-12-01& 16$\times$30       & 0.122 $\pm$ 0.007 & 51.0 $\pm$ 1.7 & yes & \\
            &     &      &      &10-12-03& 4$\times$60,16$\times$90  & 0.262 $\pm$ 0.003 & 48.2 $\pm$ 0.3 & yes & \\
            &     &      &      &11-12-03& 20$\times$60       & 0.335 $\pm$ 0.003 & 52.1 $\pm$ 0.3 & yes & \\
            &     &      &      &13-12-03& 4$\times$120,12$\times$90 & 0.288 $\pm$ 0.003 & 47.5 $\pm$ 0.3 & yes & \\
UX Ori      & 9.6 & A4   & HAe  &27-12-01& 4$\times$60        & 1.825 $\pm$ 0.063 &112.4 $\pm$ 1.0 & no  & \\
BF Ori      &10.3 & A5   & HAe  &27-12-01& 16$\times$180      & 0.886 $\pm$ 0.017 & 52.0 $\pm$ 0.6 & no  & \\       
LKha 215    &10.6 & B7.5 & HAe  &27-12-01& 16$\times$120      & 1.343 $\pm$ 0.018 & 75.9 $\pm$ 0.4 & no  & \\
HD 141569   &7.0  & B9.5 & HAe  &27-12-01& 16$\times$120      & 0.647 $\pm$ 0.005 & 85.2 $\pm$ 0.2 & yes & \\
GU Cma      &6.6  & B2   & HBe  &10-12-03& 4$\times$60        & 1.726 $\pm$ 0.006 & 27.0 $\pm$ 0.1 & yes & \\
            &     &      &      &12-12-03& 4$\times$20,8$\times$90   & 1.293 $\pm$ 0.011 & 67.6 $\pm$ 0.2 & yes & \\
MWC 166     & 7.0 & B0   & HBe  &11-12-03& 4$\times$20,12$\times$120 & 0.164 $\pm$ 0.003 & 36.8 $\pm$ 0.4 & yes & \\
            &     &      &      &12-12-03& 4$\times$120       & 0.394 $\pm$ 0.004 & 35.9 $\pm$ 0.3 & no  & \\
SV Cep      &10.1 & A2   & HAe  &12-12-03& 4$\times$60,8$\times$180  & 1.293 $\pm$ 0.011 & 67.6 $\pm$ 0.2 & yes & \\
MWC 361     & 7.4 & B2   & HBe  &13-12-03& 24$\times$30       & 0.801 $\pm$ 0.003 & 93.6 $\pm$ 0.1 & yes & (90\degree)\\
\hline
\end{tabular}
\\
\noindent
\end{minipage}
\end{table*}

\begin{table*}
\begin{minipage}{\linewidth}
\renewcommand{\thefootnote}{\thempfootnote}
\caption{The new Herbig Ae/Be star \ha\ line results. 
The errors on the equivalent widths of the \ha\ lines (column 2) 
are below 5\%, the errors on $\Delta \lambda(\rm Pol)$ (column 5) 
and $\Delta \lambda(I)$ are determined at Full Width Zero Intensity (FWZI) 
and are about 10\%.
$\Delta \lambda(\rm Pol)$ has been defined as the width over 
which the polarisation changes. In the case where the widths in PA 
and \%Pol are unequal, we take the largest of the two.  
The fractional width \ratpol\ is given in Column (6).  
The followed recipe with regard to the depolarisation question (column 7) 
is described in the text. Column (8) represents the morphology in $QU$ 
space. 
\label{t_lineh} }
\begin{tabular}{lcccccclc}
\hline
Object & \ha\ EW(\ang) & Line/Cont &  Line  & $\Delta \lambda(\rm Pol)$ & ($\frac{\Delta \lambda(\rm Pol)}{\Delta \lambda(I)}$) & depolarisation? & $QU$ & Mean\\
       &  & contrast  & effect? & (\ang) & & & behaviour & character\\
\smallskip\\
  (1) & (2)  & (3)  & (4)  & (5) & (6) & (7) & (8) & (9) \\
\hline
MWC 480     & $-$21 & 5.2  & Yes &     &      & No & exc/loop?& McLean   \\
            & $-$20 & 5.1  & Yes?&     &      & No & exc+exc?&    \\
            & $-$22 & 5.1  & Yes &     &      & No &  --    &    \\
            & $-$22 & 5.1  & Yes &     &      & No &  --    &    \\
            & $-$19 & 4.6  & Yes?&     &      & No &  --    &    \\
            & $-$19 & 4.3  & Yes?&     &      & No &  --    &    \\
AB Aur      & $-$33 & 7.3  & No  & --  & --   & -- & --     & Loop/McLean    \\
            & $-$34 & 6.9  & No  & --  & --   & -- & --     &    \\
            & $-$34 & 7.0  & Yes?&     &      & No & loop?  &    \\
            & $-$35 & 6.9  & Yes?&     &      & No & loop   &    \\
            & $-$34 & 6.9  & Yes?&     &      & No & --     &    \\
            & $-$30 & 6.1  & Yes?&     &      & No & --     &    \\
HD 141569   & $-$5.5& 1.4  & No  & --  & --   & -- & --     & None   \\
GU Cma      & $-$9.3& 2.2  & No  & --  & --   & -- & --     & None   \\
            & $-$9.3&      &     &     &      & No & --     &    \\
MWC 166     & $+$0.59&0.75 & No  & --  & --   & -- & --     & Abs   \\
SV Cep      & $-$11 & 2.0  & Yes?& 20  & 1.0  & No & loop   & Loop   \\
MWC 361     & $-$47 & 8.2  & Yes &     &      & Yes($+$)& exc/loop? & Depol+Loop \\
\hline
\end{tabular}
\\
\end{minipage}
\end{table*}

\subsection{Complementary Herbig Ae/Be star data}

\subsubsection{MWC~480}

MWC~480 was the first Herbig Ae star for which a rotating structure has been detected at longer wavelengths. 
Mannings et al. (1997) found an inclined disk, $i$ = 30\degree, with a PA = 157\degree\ $\pm$ 4\degree.
Simon et al. (2000) derive a disk PA and inclination 
of respectively PA = 148\degree\ $\pm$ 1\degree, $i$ = 38\degree\ $\pm$ 1\degree.
Augereau et al. (2001) do not detect the disk with HST, although they argue for disk rotation over an angular 
scale of a few arcsec.
Eisner et al. (2004) find an inclined disk with  PA $\simeq$ 150\degree, $i$ $\simeq$ 30\degree.
I.e. all imaging studies agree on a PA of $\simeq$ 150\degree.\\

{\it Our data:~}
Since we did not detect significant variability in the line polarisation spectra of MWC~480, taken
in a few nights of 2001 December, we have combined these data, and obtained a higher S/N.
The combined data are shown in Fig.~\ref{f_yesherbig}. 
In accordance with the data in Vink et al. (2002), there is an excursion present in the $QU$ plane, 
consistent with the McLean effect, but there is 
no detection of a $QU$ loop.
The PA of the $QU$ plane excursion is 60\degree, which is consistent with a 90-degree 
rotation compared to the above-mentioned imaging and interferometry studies.

\subsubsection{MWC 361 = HD 200775}

The star is known to be a binary (Li et al 1994; see also Millan-Gabet, Schloerb \& Traub 2001), 
with MWC~361~B positioned 
at a PA of 164\degree\ (see Pirzkal, Spillar \& Dyck 1997). Pogodin et al. (2004) present high-resolution spectroscopy to 
search for periodicity, but these attempts have so far been unsuccessful.
Miroshnichenko et al. (1998) report high \ha\ emission activity, which they claim is recurrent on a 3.68 year timescale.\\

{\it Our data:~}
We notice complex changes across \ha\ in the polarisation triplot, accompanied by a complex structure in the $QU$ plane.
In Vink et al. (2002) we noted the possibility of the presence of a linear excursion and a loop, which 
is here confirmed on the basis of higher quality data.

\subsubsection{SV Sep}

We show the data of SV~Cep because of the well-observed loop in the $QU$ plane, the loop 
is more pronounced than in Vink et al. (2002).

\subsubsection{HD~141569}

HD~141569 is known to be an object that is transitional between a PMS star and 
a main sequence star with a debris disk.
HST coronographic observations by Augereau et al. (1999) revealed a dust ring that peaks at 
approximately 325 AU. This disk is found to be inclined at 37.5\degree\ $\pm$ 4.5\degree\ to the line 
of sight (see also Weinberger et al. 1999; Mouillet et al. 2001).  
The disk appears to have a central gap extending out to 17 -- 30 AU (Sylvester \& Skinner 1996; 
Marsh et al. 2002; Brittain et al. 2003). \\

{\it Our data:~}
The polarisation data of HD~141569, shown in Fig.~\ref{f_noherbig}, do not indicate a line effect. 
This could possibly be due to a lack of scattering particles, caused by the large central gap. 

\subsubsection{MWC~166~(=~HD~53367) \& GU~CMa~(=~HD~52721)}

MWC~166 has shown significant changes in its line depolarisation character, which we have linked 
to changes in the equivalent width (see Vink et al. 2002). Here, the \ha\ line is mainly in absorption. 
Such dramatic changes in the \ha\ equivalent width are not uncommon for classical Be stars 
(cf. the extreme case of HD~76534 Oudmaijer \& Drew 1997)
For completeness, the other Type III (according to Hillenbrand et al. 1992) Herbig Be star that we 
have re-observed, GU~Cma, has not yet shown depolarisations.

\subsubsection{AB~Aur  --  not plotted}

AB~Aur is the brightest Herbig star in the sky and a member of the ``B8 -- A2 subclass'' 
where P Cygni absorption is more often observed. 
We have already shown in Vink et al. (2002) that this is often accompanied by the McLean effect. 

Mannings \& Sargent (1997) have detected a rotating disk of 450 AU in the $^{13}$CO (J = 1-0) line, and 
they derived an inclination of 76\degree\ and a PA of 79\degree.
However, Grady et al. (1999) found an upper limit to the inclination of 45\degree\ with HST. 
Eisner et al. (2003) found an inclination $i$ = 27\degree\ -- 35\degree, and revised 
this to $i$ = 8\degree\ -- 16\degree\ in Eisner et al. (2004).
Fukagawa et al. (2004) have recently performed coronographic H-band imaging, and these authors 
derive a disk PA and inclination of respectively PA = 58\degree\ $\pm$ 5\degree\ and 
$i$ = 30\degree\ $\pm$ 5\degree.
The apparent inconsistency between the inclination derived from the mm studies and the 
lower wavelength studies may raise the question of whether the AB~Aur disk could be significantly 
warped. However, recent kinematic modelling of the outer mm disk seems to suggest that 
the inclination is $\la$ 30 degrees (see Natta et al. 2001; Corder, Eisner \& Sargent 2005). \\

{\it Our data:~}
From the McLean effect in Vink et al. (2002) we find an intrinsic PA of 160\degree.
This is in good agreement with a 90-degree rotation of the PA of the polarisation 
compared to all above-mentioned imaging and interferometry studies.
If we adopt the likely inclination of $\simeq$ 30\degree, we can use the Vink et al. (2005) 
$i$ -- $\rin$ table to set a lower limit to the disk inner hole of AB~Aur from 
the presence of the single loop in the $QU$ plane: $\rin$ $\ga$ 5 $\rstar$. 

\subsection{A summary of Herbig Ae/Be star PAs}

The complementary Herbig Ae/Be star data have not changed the 
statistics, nor challenged the conclusions reached in Vink et al. (2002). 
However, since new imaging results with PA and $i$ information have recently become 
available, we have summarised the new information in the form of Table~\ref{t_pa}.
We have only included sources for which we can infer a well-defined PA
of the polarisation from the $QU$ plane, and give more uncertain values 
in brackets. 
For most of the T~Tauri and some of the Herbig Ae stars, PA information is already available 
from imaging studies, but for the more massive PMS stars such information is 
lacking, due to the larger distances involved. Nonetheless, such information may 
become available, if not spatially resolved, then maybe through interferometric analyses.

In addition to the objects studied in this paper, we draw attention to CQ~Tau.
VLA data by Testi et al. (2003) suggest the presence of an inclined disk with $i$ $\simeq$ 70\degree. 
However, Eisner et al. (2004) find the disk inclined with $i$ $\simeq$ 48\degree, 
at a PA = 104 -- 106 \degree. Kinematic modelling of more sensitive mm data 
is reported to show that also the outer disk is inclined by $\simeq$ 45\degree\ (Corder et al., in preparation).

The data reported in Vink et al. (2002) show the presence of the McLean effect across 
the {\it red}-shifted absorption. This occurs at a PA of 20\degree. 
Interestingly, this is consistent with a 90\degree\ rotation compared to the 
imaging studies.
The reported inclination of $\simeq$ 45\degree\ may subsequently again be used to constrain the disk 
inner hole of CQ Tau: $\rin$ $\ga$ 4 $\rstar$.

\section{Discussion}
\label{s_disk}

\subsection{The inner disk in PMS stars}

One of the motivations behind this study is to discover whether there
is a dependence of the circumstellar geometry of
PMS stars on stellar mass.  

\begin{figure} 
\mbox{\epsfxsize=0.49\textwidth\epsfbox{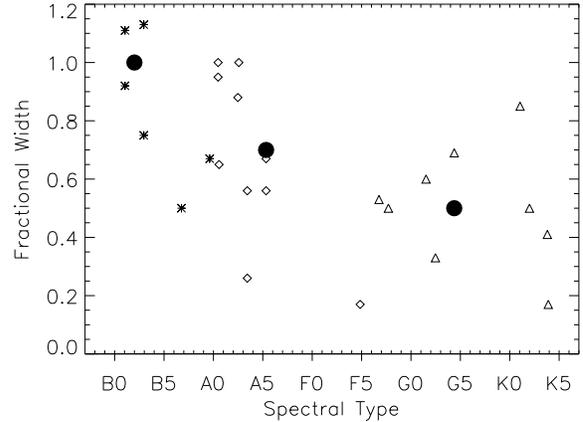}}
\caption{The fractional width \ratpol\ plotted against spectral 
type. The stars indicate the Herbig Be stars, the open diamonds represent the Herbig Ae stars, 
whilst the T~Tauri stars are indicated by open triangles.
The errors on the fractional width \ratpol\ are about 
10 per cent. The black circles indicate the averages of the subgroups as discussed in the text.}
\label{f_trend}
\end{figure}

First we restate the raw statistics of the frequency of detectable 
line effects across \ha\ among the Herbig Ae/Be and T Tauri stars.
We have previously found that about half of
the HBe stars have detectable changes across the line (7 out of 12 to be 
precise), while this rate rises to 9 out 
of 11 for the HAe stars, and here we find that this fraction is 9/10 for T Tauri stars. 
In Vink et al. (2002) we showed that in the case of Herbig Ae/Be stars, the fractional 
widths tend to decrease toward later spectral type (see also Fig.~\ref{f_trend}).
To study whether there is any further dependence on spectral 
type for T Tauri stars, we have combined the fractional widths with the Herbig data from 
Vink et al. (2002) and plotted them in Fig.~\ref{f_trend}.  
There is much scatter, partly due to the difficulty of
measurement and uncertainties in spectral typing, but presumably also due to real 
physical scatter.
Nonetheless, it is apparent that the fractional widths for e.g. the B0-B2
HBe stars are higher than those measured at spectral types later
than A2. 
Vink et al. (2002) found that \ratpol\ = 1.0 $\pm$ 0.2 
for the early Herbig Be (B0 -- B2) group.
With the new Herbig Ae star data, we have updated the fractional width for the Herbig Ae group 
to \ratpol\ = 0.7 $\pm$ 0.3. Finally, for the T Tauri stars, 
we find \ratpol\ = 0.5 $\pm$ 0.2.
%, whilst this value was 0.5 $\pm$ 0.3 
%for the late Herbig Ae group (later than A2). 
%Here we find 0.5 $\pm$ 0.2 for the T Tauri stars, which is similar to the
%late Herbig Ae stars, and smaller than the Herbig Ae group as a whole, which 
%have an average \ratpol\ value of 0.7 $\pm$ 0.3. 
%%Rather than statistics alone, 

Of further significance is that the late Herbig Ae and 
T~Tauri stars present similar characteristics in their \ha\ line polarimetry, and this 
stands in sharp contrast to the \ha\ depolarisations that are  
seen in the Herbig Be stars. The depolarisations simply signal a 
flattened geometry, consistent with an undisrupted disk on small scales, but kinematic 
information on the inner gas is not obtained.
% -- the situation is very similar to that of classical Be stars. 
On the other hand, the $QU$ loops found 
for the Herbig Ae and T~Tauri stars are due to intrinsic {\it line} polarisation that is 
caused by scattering of a compact source of \ha\ photons off an exterior rotating disk. 
It is tempting to associate the presence of a compact source of \ha\ photons and a rotating undisrupted disk 
with the magnetospheric accretion scenario, but this is not a required interpretation. Nevertheless, 
the observed single $QU$ loops, as for instance measured in the magnetically  
active T~Tauri star BP~Tau and many other T Tauri and  
Herbig Ae stars, are consistent with the properties expected of magnetospheric accretion -- a model that was proposed to 
explain a range of phenomena other than those reported here.
%may be regarded as a viable signpost of magnetic accretion, as the stellar 
%magnetic field would indeed be expected to disrupt the inner disk up to a few stellar radii.}
%From our data, we deduce the presence of rotating undisrupted disks in T Tauri and Herbig Ae stars, but we 
%stress that this inference does not depend on the adoption of the magnetic accretion scenario.} 

\begin{table}
\begin{minipage}{\linewidth}
\renewcommand{\thefootnote}{\thempfootnote}
\caption{Comparison (column 5) between intrinsic position angles of the polarisation (column 3) 
and those from imaging studies (column 4) for Herbig Ae/Be stars (above) and T Tauri stars (below). 
The polarisation PA data is taken from Vink et al. (2002,2003) and the present study. 
Uncertain PA data is given in brackets. Note the lack of imaged disk PAs, especially for the Herbig stars.}
\label{t_pa}
\begin{tabular}{llrrr}
\hline
Name & Spec. & Polarisation  & Imaging                     & $\Delta$ PA     \\
     & Type  & PA   & PA &              \\
\hline
MWC 137 & Be    &  25  &   & \\
MWC 166 & B0    &  50  &   & \\
MWC 361 & B2    & (90) &   & \\
BD$+$40 4124 & B2 &  83  &   & \\
HD~45677 & B3   &$\simeq$70 & & \\
MWC~158  & B9   & 135  &   & \\
HD~58647 & B9   &  20  &   & \\
AS~477   &B9.5   & 110  &   & \\
MWC~120  & A0   & 90   &   & \\
KMS 27   & A0   & (65) &   & \\
MWC~789  & A0   & 175  &   & \\
AB Aur   & A0   & 160  & 60-80$^2$ & 80 -- 100\\
SV Cep   & A2   & (100)&   & \\
XY Per   & A2   & (70) &   & \\
HD~244604 & A3   & $\simeq$155  &   & \\
T Ori    & A3   & 20   &   & \\
HD~245185 & A5  & $\simeq$40 & \\
MWC 480 & A5    &  55  & 150$^1$  & 95\\
CQ~Tau  & F2    & 20   & 105$^3$  & 95 \\
\hline
CO Ori  & F7    & (60) &     & \\
RY Tau  & F8    & 163  & 62$^4$  & 101\\
\\
SU Aur  & G2    & 130  & 127$^5$  & 3\\ 
FU Ori  & G3    &  45  &  47$^6$  & 2\\
GW Ori  & G5    & (60) & 56$^8$? & (4?)\\
UX Tau A & K2   & 140  &     & \\
RW Aur A & K3   & 115  &     & \\
DR Tau   & K5   & 120  & 128$^7$  & 8\\
\hline
\end{tabular}
\\
\noindent
$^1$ Eisner et al. (2004), Mannings et al. (1997), Simon et al. (2000)
$^2$ Fukagawa et al. (2004), Corder et al. (2005)
$^3$ Eisner et al. (2004)
$^4$ Akeson et al. (2003), Koerner \& Sargent (1995), Kitamura et al. (2002)
$^5$ Akeson et al. (2002)
$^6$ Malbet et al. (2005)
$^7$ Kitamura et al. (2002)
$^8$ Mathieu et al. (1995)
\end{minipage}
\end{table}

From our linear spectropolarimetry data we find additional results which 
would be hard to obtain from straightforward spectroscopy.
In a few cases, most notably FU~Ori and DR~Tau, we find the McLean effect: an increase in the degree
of polarisation across Stokes $I$ absorption, usually seen blueshifted, but it may also be observed 
in the red wing, cf. CQ Tau. Modeling of these effects may become a worthwhile exercise, given 
their frequent discovery in strongly accreting objects such as FU~Ori. 
The extra information from the spectropolarimetry may provide vital constraints for 
understanding disk winds in young stellar objects. 
Furthermore, the McLean polarisation effects give a particularly accurate value for the intrinsic 
PA of the polarisation. 

We have therefore searched the literature for information on position angles 
and inclinations of known imaged disks by infrared/millimetre imaging and interferometry. 
Although there is a lack of imaged disk PAs, there is consistency between the PAs found from our \ha\ line 
polarimetry and the larger disks from imaging studies, cf. column (4) in Table~\ref{t_pa}. 
For the three Herbig Ae stars with imaged disk PAs, 
our polarisation PAs are perpendicular to these larger disk PAs, consistent with the picture of 
an optically thin inner gaseous disk for Herbig Ae stars . A similar result is found for 
the F8 T Tauri star RY~Tau, but for the later-type T Tauri stars and FU Ori, the imaged 
disk PAs are aligned with the polarisation PAs -- indicating optically thick, multiple scattering inner disks. 
Although we find the geometry and kinematics of T Tauri and Herbig Ae stars to be very similar, i.e. 
rotating inner disks, the optical depths of these disks appear to be different. There seems to be 
a change from optically thick disks in T Tauri stars to optically thin ones for Herbig Ae stars. 
%In particular, the T Tauri/FU Orionis-type objects appear to posses ``standard'' Shakura \& Sunyaev optically 
%thick accretion disks, whereas the Herbig Ae disks appear to be optically thin. 
This appears to be consistent with recent NIR interferometry results (e.g. Monnier et al. 2005). 
The optically thin disk in Herbig Ae stars may allow the stellar radiation to hit the inner wall of the outer 
dust disk (located at roughly 1 AU), and to ``puff it up'' as described by Dullemond, Dominik \& Natta (2001).

%{\bf Furthermore, we repeat the potential of our novel technique to determine 
%the size of the inner holes of rotating disks.}
%Since the line polarimetry models of Vink et al. (2005) set 
%combined constraints on disk inclination and disk inner hole, the literature 
%inclinations were used to set constraints on the disk inner hole sizes of the 
%objects SU~Aur, GW~Ori, AB~Aur, and CQ~Tau. 

\subsection{The connection with the orientation of the magnetic 
field in the foreground material}

In the basic picture of the star formation process (e.g. Uchida \& Shibata 1985), one 
would expect the stellar accretion disk axis to be parallel to the global magnetic field. 
Although there are still many uncertainties in  
the physics of grain alignment (e.g. Roberge \& Lazarian 1999), one would normally  
expect the grains to align perpendicular to the global magnetic field. This would yield 
foreground polarisation with the observed polarisation PA parallel to this global field. 
Imaged disks should hence be found perpendicular to the direction of foreground polarisation. At first 
this was confirmed (e.g. Tamura \& Sato 1988), but more recently it has been called into 
question by M\'enard \& Duchene (2004), who found the disk orientations in Taurus to be 
consistent with a random orientation. We have seen in Sect.~\ref{s_fore} that for those cases where we 
can infer the PAs of the intrinsic {\it and} the foreground 
polarisation, these are 90-degrees rotated from each other.
Although the cases where we are able to derive these intrinsic and foreground 
PAs simultaneously are few at the present time, these data nonetheless hint at 
an undisturbed star-forming collapse where the orientation of the environment's magnetic 
field is preserved. If dynamical effects (e.g. Bate, Bonnell \& Bromm 2003) 
are important in nearby star-forming molecular clouds such as Taurus and Orion, this would typically  
break the connection between the stellar axis and the foreground magnetic field direction -- in contradiction 
to the results currently at hand.
%have a disruptive influence on this symmetry. In other words, the connection between the stellar 
%magnetic field and that of the local environment would be broken -- contradictory to what we find.}  
%that proceeds ``statically'' along global magnetic fields, without 
%the need for a more dynamical star-formation process (e.g. Bate, Bonnell \& Bromm 2003) in nearby 
%star-forming regions such as Taurus-Auriga and Orion. 

\section{Conclusions}
\label{s_con}

We have presented \ha\ spectropolarimetry observations of a sample of 10 bright 
T Tauri stars, supplemented with new Herbig Ae/Be star data. From these data we draw the 
following conclusions:\\

\begin{itemize}

\item{} The changes in the linear polarisation across \ha\ in T~Tauri (9/10) and 
Herbig Ae (9/11) objects are consistent with the presence of a compact source of line emission  
that is scattered off a rotating inner accretion disk.\\

\item{} We find consistency between the PA of the 
polarisation and those of known disk PAs from infrared and millimetre imaging 
and interferometry studies, that measure larger scales.\\

\item{} For the Herbig Ae stars, AB~Aur, MWC~480 and CQ~Tau, and the early-type T Tauri star 
RY~Tau, we find the PA of the polarisation to be perpendicular to the imaged disk -- indicative 
of optically thin inner disks for Herbig Ae stars.\\

\item{} The polarisation PA is found to be parallel with the imaged disk 
for the T Tauri stars DR~Tau and SU~Aur, as well as FU~Ori. This is consistent with 
multiple scatterings in optically thick accretion disks.\\

\item{} We can constrain the sizes of disk inner holes by combining known inclination angles with 
the results of Monte Carlo scattering models for the disks of SU~Aur ($\Rin$ $>$ 3 $\Rstar$), 
GW~Ori ($\Rin$ $\simeq$ 3--4 $\Rstar$), 
AB~Aur ($\Rin$ $\ga$ 5 $\Rstar$), 
and CQ~Tau ($\Rin$ $\ga$ 4 $\Rstar$).\\

\item{} For FU~Ori and DR~Tau, we infer the PA of the intrinsic and foreground polarisations to 
be mutually perpendicular. This suggests an undisturbed star-forming collapse in which 
a ``memory'' of the orientation of the large scale magnetic field is preserved.\\

\end{itemize}

\paragraph*{\it Acknowledgments}

We thank the referee for comments that have improved 
the content of this paper. The allocation of time on the William Herschel 
Telescope was awarded by PATT, the United Kingdom allocation panel.  
JSV is funded by the PPARC.
The data analysis facilities are provided by the Starlink Project, which 
is run by CCLRC on behalf of PPARC. This research has made use of the
{\sc simbad} database, operated at CDS, Strasbourg, France.


\begin{thebibliography}{}

\bibitem[]{}
Akeson R.L., Koerner D.W., Jensen W.L.N., 1998, ApJ 505, 358

\bibitem[]{}
Akeson R.L., Ciardi D.R., van Belle G.T., Creech-Eakman M.J., 2002, ApJ 566, 1124

\bibitem[]{}
Akeson R.L., Ciardi D.R., van Belle G.T., 2003, SPIE 4838, 1037

\bibitem[]{}
Alencar S.H.P., Johns-Krull C.M., Basri G., 2001, AJ 122, 3335

\bibitem[]{} 
Angel J.R.P., 1969, ApJ 158, 219

\bibitem[]{}
Artymowicz P., Lubow S.H., 1994, ApJ 421, 651

\bibitem[]{}
Augereau J.C., Lagrange A.M., Mouillet D., M\'enard F., 1999, A\&A 350, 51

\bibitem[]{}
Augereau J.C., Lagrange A.M., Mouillet D., M\'enard F., 2001, A\&A 365, 78

\bibitem[]{}
Basri G., Martin E.L., Bertout C., 1991, A\&A 252, 625

\bibitem[]{}
Bastien P., 1982, A\&AS 48, 153

\bibitem[]{} 
Bastien P., 1985, ApJS 59, 277

\bibitem[]{} 
Bastien P., Landstreet J.D., 1979, ApJ 229, 137

\bibitem[]{}
Bastien P., M\'enard, 1988 ApJ 326, 334

\bibitem[]{}
Bate M.R., Bonnell I.A., Bromm V., 2003, MNRAS 339, 577

\bibitem[]{}
Bergner Yu. K., Miroshnichenko A.S., Yudin R.V., Yu N., Yutanov K.G., Dzhakusheva K.G., Mukanov D.B., 1987, PAZh 13, 208

\bibitem[]{}
Bouvier J., \& Bertout C., 1989, A\&A 211, 99

\bibitem[]{}
Bouvier J., Bertout C., Benz W., Mayor M., 1986, A\&A 165, 110

\bibitem[]{}
Brittain S.D., Rettig T.W., Simon T., Kulesa C., DiSanti M.A., Dello Russo N., 2003, ApJ 588, 535

\bibitem[]{}
Brown J.C., McLean I.S., 1977, A\&A 57, 141

\bibitem[]{}
Calvet N., Hartmann L., Kenyon S.J., Whitney B.A., 1994, ApJ 434, 330

\bibitem[]{}
Catala C., et al., 1999, A\&A 345, 884

\bibitem[]{}
Clarke D., McLean I.S. 1974, MNRAS 167, 27

\bibitem[]{}
Corder S., Eisner J., Sargent A., ApJL, in press, astro-ph/0502131

\bibitem[]{}
DeWarf L.E., Sepinsky J.F., Guinan E.F., Ribas I., Nadalin I., 2003, ApJ 590, 357

\bibitem[]{}
Drissen L., Bastien P., St-Louis N., 1989, AJ 97, 814

\bibitem[]{}
Dougherty S.M., Taylor A.R., 1992, Nature 359, 808

\bibitem[]{}
Dullemond C.P., Dominik C., Natta A., 2001, ApJ 560, 957

\bibitem[]{}
Dutrey A., Guilloteau S., Simon M., 2003, A\&A 402, 1003

\bibitem[]{}
Dyck H.M., Simon T., Zuckerman B., 1982, ApJ 225, 103

\bibitem[]{}
Eisner J.A., Lane B.F., Akeson R.L., Hillenbrand L.A., Sargent A.I., 2003, ApJ 588, 360

\bibitem[]{}
Eisner J.A., Lane B.F., Hillenbrand L.A., Akeson R.L., Sargent A.I., 2004, ApJ 613, 1049

\bibitem[]{}
Folha D.F.M., Emerson J.P., 1999, A\&A 352, 517 

\bibitem[]{}
Fukagawa M., 2004, ApJ 605, 53

\bibitem[]{}
Gahm G.F., Gullbring E., Fischerstrom C., Lindroos K.P., Loden K., 1993 A\&AS 100, 317

\bibitem[]{}
Ghez A.M., Neugebauer G., Gorham P.W., Haniff C.A., Kulkarni S.R., Matthews K., Koresko C., Beckwith S., 1991, AJ 102, 2066

\bibitem[]{}
Ghez A.M., Neugebauer G., Matthews K., 2003, AJ 106, 2005

\bibitem[]{}
G\'omez de Castro A.I., Verdugo E., 2003, ApJ 597, 443

\bibitem[]{}
Grady C.A., Woodgate B., Bruhweiler F.C., Boggess A., Plait P., Lindler D.J., Clampin M., Kalas P., 1999, ApJ 523, L151

\bibitem[]{}
Greaves J.S., 2004, MNRAS 351 99

\bibitem[]{}
Grinin V.P., Kolotilov E.A., Rostopchina A., 1995, A\&AS 112, 457

\bibitem[]{}
Gullbring E., \& Gahm G.F., 1986, A\&A 308 821

\bibitem[]{} 
Hartmann L., 1999, NewAR 43, 1 

\bibitem[]{}
Hartmann L., Kenyon S.J., 1985, ApJ 299, 462

\bibitem[]{}
Hartmann L., Hewett R., Stahler S., Mathieu R.D., 1986, ApJ 309, 275

\bibitem[]{}
Heines A., Henning T., Szeifert T., 1997, IAUS 182, 294

\bibitem[]{}
Herbig G.H., 1977, ApJ 217, 693

\bibitem[]{}
Herbig G.H., Bell R.K., 1988, Lick Observatory Bulletin, 1111

\bibitem[]{}
Herbst W., et al., 1987, AJ 94, 137

\bibitem[]{}
Hillenbrand L.A., Strom S.E., Vrba F.J., Keene J. 1992, ApJ 397, 613

\bibitem[]{}
Hogerheijde M.R., van Langevelde H.J., Mundy L.G., Blake G.A., van Dishoeck E.F., 1997, ApJ 490, 99

\bibitem[]{} 
Hough J.H., Bailey J., Cunningham E.C., McCall A., Axon D.J., 1981, MNRAS 195, 429

\bibitem[]{}
Johns C.M., Basri G., 1995, AJ 109 2800

\bibitem[]{}
Johns-Krull C.M., Valenti J.A., Koresko C., 
1999, ApJ 516, 900 

\bibitem[]{}
Kitamura Y., Momose M., Yokogawa S., Kawabe R., Tamura M., Ida S., 2002, ApJ 581, 357
 
\bibitem[]{}
Koerner D.W., Sargent A.I., 1995, AJ 109, 2138

\bibitem[]{}
Koresko C.D., 2000, ApJ 531, 147

\bibitem[]{}
Kravtsova A.S., Lamzin S.A., 2002, AstL 28, 676

\bibitem[]{}
Lamzin S.A., Shevchenko V., Grankin K., Mel'nikov S., 1998, Ap\&SS 261, 167

\bibitem[]{}
Li W., Evans N.J., Harvey P.M., Colome C., 1994, ApJ 433, 199
 
\bibitem[]{}
L\'opez-Mart\'in L., Cabrit S., Dougados C., 2003, A\&A 405

\bibitem[]{}
Malbet F., et al., 1998, ApJ 507, 149

\bibitem[]{}
Malbet F., et al., 2005, A\&A submitted

\bibitem[]{}
Mannings V., Sargent A.I., 1997, ApJ 490, 792

\bibitem[]{}
Mannings V., Koerner D.W., Sargent A.I., 1997, Nature 388, 555

\bibitem[]{}
Marsh, K.A., Silverstone M.D., Becklin E.E., Koerner D.W., Werner M.W., Weinberger A.J., Ressler M.E., 2002, ApJ 573, 425

\bibitem[]{}
Mathieu R.D., Adams F.C., Latham D.W., 1991, AJ 101, 2184

\bibitem[]{}
Mathieu R.D., Adams F.C., Fuller G.A., Jensen E.L.N., Koerner D.W., Sargent A.I., 1995, AJ 109, 2655

\bibitem[]{}
McKee C.F., Tan J.C., 2003, ApJ 585, 850
  
\bibitem[]{}
McLean I.S., 1979, MNRAS 186, 265

\bibitem[]{}
M{\'e}nard F., Bastien P., 1992, AJ 103, 564

\bibitem[]{}
M\'enard F., Duch\^ene G., 2004, A\&A 425, 973

\bibitem[]{} 
Millan-Gabet R., Schloerb F.P., Traub W.A., 2001, ApJ 546, 358

\bibitem[]{}
Miroshnichenko A.S., Mulliss C.L., Bjorkman K.S., Morrison N.D., Glagolevskij Y.V., Chountonov G.A., 1998, PASP 110, 883

\bibitem[]{}
Monnier J.D., Millan-Gabet R., Billmeier R., Akeson R., Wallace D., et al., 2005, ApJ in press, astro-ph/0502252

\bibitem[]{}
Mora A., et al., 2001, A\&A 378, 116

\bibitem[]{}
Mouillet D., Lagrange A.M., Augereau J.C., M\'enard, F., 2001, A\&A 372, 61

\bibitem[]{}
Mundt R., Walter F.M., Feigelson E.D., Finkenzeller U., Herbig G.H., Odell A.P., 1983, ApJ 269, 229

\bibitem[]{}
Murdin P., Penston M.V., 1977, MNRAS 181, 657

\bibitem[]{}
Muzerolle J., Calvet N., Hartmann L., 2001, ApJ 550, 944

\bibitem[]{}
Najita J., Carr J.S., Mathieu R.D., 2003, ApJ 589, 931

\bibitem[]{}
Natta A., Prusti T., Neri R., Wooden D., Grinin V.P., Mannings V., 2001, A\&A 371, 186

\bibitem[]{}
Oudmaijer R.D., Drew J.E., 1997, A\&A 318, 198

\bibitem[]{}
Oudmaijer R.D., Drew J.E. 1999, MNRAS 305, 166

\bibitem[]{}
Oudmaijer R.D., et al., 2001, A\&A 379, 564

\bibitem[]{}
Petrov P.P., Gahm G.F., Gameiro J.F., Duemmler R., Ilyin I.V., Laakkonen T., Lago M.T.V.T., Tuominen I., 2001, A\&A 369, 993

\bibitem[]{}
Pirzkal N., Spillar E.J., Dyck H.M., 1997, ApJ 481, 392	

\bibitem[]{}
Poeckert R. 1975, ApJ 152, 181

\bibitem[]{}
Poeckert R., Marlborough J.M., 1976, ApJ 206, 182

\bibitem[]{}
Pogodin M.A., et al., 2004, A\&A 417, 715	
	
\bibitem[]{}
Quirrenbach A., et al., 1997, ApJ 479, 477

\bibitem[]{}
Roberge W.G., Lazarian A., 1999, MNRAS 305, 615

\bibitem[]{}
Rucinski S.M., 1985, AJ 90, 2321

\bibitem[]{}
Schulte-Ladbeck R.E., 1983, A\&A 120, 203

\bibitem[]{}
Serkowski K., 1969, ApJ 156, 55

\bibitem[]{}
Serkowski K., Mathewson D.L., Foid V.L., 1975, ApJ 196, 261

\bibitem[]{}
Shevchenko V., Grankin K.N., Mel'Nikov S.Y., Lamzin S.A., 1998, PAZh, 24, 614

\bibitem[]{}
Simon M., Dutrey A., Guilloteau S., 2000, ApJ 545, 1034

\bibitem[]{}
Smirnov D.A., Fabrika S.N., Lamzin S.A., Valyavin G.G., 2003, A\&A 401, 1057

\bibitem[]{}
Smith K.W., Lewis G.F., Bonnell I.A., Bunclark P.S., Emerson J.P., 1999, MNRAS 304, 367
 
\bibitem[]{}
Stassun K., Wood K., 1999, ApJ 510, 892

\bibitem[]{}
Sylvester R.J., Skinner C.J., 1996, MNRAS 283, 457

\bibitem[]{}
Symington N.H., Harries T.J., Kurosawa R., Naylor T., 2005, MNRAS, in press

\bibitem[]{}
Tamura M., Sato S., 1988, AJ 98, 1368

\bibitem[]{}
Testi L., Natta, A., Shepherd D.S., Wilner D.J., 2003, A\&A 403, 323

\bibitem[]{}
Uchida Y., Shibata K., 1985, PASJ 37, 515

\bibitem[]{}
Unruh Y.C., et al., 2004, MNRAS 348, 1301

\bibitem[]{}
Vardanyan R.A., 1964, SoByu 35, 3

\bibitem[]{} 
Vink J.S., Drew J.E., Harries T.J., Oudmaijer R.D., 2002, MNRAS 337, 356

\bibitem[]{}
Vink J.S., Drew J.E., Harries T.J., Oudmaijer R.D., Unruh Y.C., 2003, A\&A 406, 703

\bibitem[]{}
Vink J.S., Harries T.J., Drew J.E., 2005, A\&A 430, 213

\bibitem[]{}
Wang H., Apai D., Henning T., Pascucci I., 2004, ApJ 601, 83

\bibitem[]{}
Weinberger A.J., Becklin E.E., Schneider G., Smith B.A., Lowrance P.J., Silverstone M.D., Zuckerman B., Terrile R.J., 1999, ApJ 525, 53

\bibitem[]{}
Wood K., Brown J.C., Fox G.K., 1993, A\&A 271, 492

\bibitem[]{}
Wood K., Bjorkman K.S., Bjorkman J.E., 1997, ApJ 477, 926

\bibitem[]{}
Yudin R.V., 2000, A\&AS 144, 285

\end{thebibliography}
\end{document}